\documentclass[12pt]{article}
\usepackage{hyperref}
\usepackage{multicol}
\usepackage{bm}
\usepackage{hyperref}
\usepackage{graphicx}  
\usepackage{dcolumn}   
\usepackage{bm}        
\usepackage{amssymb}   
\usepackage{amsmath}
\usepackage{amsfonts}
\usepackage[usenames]{color}

\newcommand{\GeV}      {~\mathrm{GeV}}
\newcommand{\pb}      {~\mathrm{pb}}

\def\co{coannihilation~}

\def\pt{\not\!\!{P_T}}

\def \chan{\widetilde{\chi}}
\def \cha{\widetilde{\chi}^{\pm}_1}

\def \na{\widetilde{\chi}^{0}}
\def \nb{\widetilde{\chi}^{0}_2}

\def \g{\tilde{g}}

\def \ta{\widetilde{t}_1}

\def \sta{\widetilde{\tau}_1}

\def \mh{m_{1/2}}
\def\.4{\vspace{-.5cm}}

\def\beq{\begin{equation}}
\def\be{\begin{equation}}
\def\beqn{\begin{eqnarray}}
\def\ee{\end{equation}}
\def\eeq{\end{equation}}
\def\eeqn{\end{eqnarray}}

\def\a{GNLSP$_{\rm A}~$}
\def\b{GNLSP$_{\rm B}~$}
\def\c{GNLSP$_{\rm C}~$}

\begin{document}

\setlength{\baselineskip}{0.2in}

\begin{titlepage}
\noindent
\begin{flushright}
NUB-3263\\
YITP-SB-09-10
\end{flushright}

\begin{center}
  \begin{Large}
    \begin{bf}
\Huge Gluino NLSP, Dark Matter via Gluino Coannihilation, and LHC Signatures\\
     \end{bf}
  \end{Large}
\end{center}
\vspace{0.2cm}

\begin{center}

\begin{large}
Daniel Feldman$^1$, Zuowei Liu$^{2}$, and Pran Nath$^1$
\end{large}
\vspace{0.55cm}
\\
  \begin{it}
$^1$Department of Physics, Northeastern University,
 Boston, MA 02115, USA,\vspace{0.25cm}
\\$^2$C. N. Yang Institute for Theoretical Physics, \\Stony Brook University, Stony Brook, NY 11794, USA.
\vspace{0.5cm}
\end{it}\\

\end{center}


\begin{abstract}
The possibility that the gluino is  the next to the lightest supersymmetric particle (NLSP) is discussed and it is
shown that this situation arises in nonuniversal SUGRA models within a significant part of the
 parameter space compatible with all known experimental bounds.
  It is  then shown that the gluino NLSP (GNLSP) models lead to a compressed sfermion spectrum
with the sleptons often heavier than the squarks at least for the first two generations.
The relic density here is governed by gluino coannihilation 
which is responsible for a relatively small mass splitting between the gluino and the neutralino masses. 
Thus the GNLSP class of models is very predictive first because the SUSY production cross
sections at the LHC are dominated by gluino production and second  because the gluino production
itself  proceeds dominantly through a single channel which allows for  a direct 
determination of the gluino mass  and an indirect determination of  the neutralino mass due
to a linear relation between 
these two masses which
is highly constrained by coannihilation.  A detailed analysis of these models shows that the 
 jet production and tagged b-jets from  the gluino production can be discriminated from 
 the standard model background with appropriate cuts.
 It is found that the GNLSP models can be tested with just 10 fb$^{-1}$ of integrated
luminosity and may therefore be checked  with low luminosity runs in the first data at the  LHC.
Thus if a GNLSP model is realized, the LHC will turn into a gluino factory through a profuse production
of gluinos with typically only a small fraction $\lesssim 5\%$  of total SUSY events arising from other production 
modes over the allowed  GNLSP model parameter space.

\end{abstract}

\end{titlepage}

\section{Introduction}
One of the interesting possibilities that arises within the landscape
of possible sparticle mass hierarchies \cite{Feldman:2007zn} 
is 
that the gluino $(\g)$ is the next to the lightest
supersymmetric particle (NLSP) where neutralino dark matter
produces the correct relic abundance of such matter consistent
with the WMAP observations \cite{Spergel:2006hy}.   
In fact, an analysis in the 
context of nonuniversal supergravity models (NUSUGRA)
reveals that the gluino NLSP model (GNLSP) arises in a significant part of the parameter space \cite{Feldman:2007fq,Feldman:2008hs}.
Amongst the various possible ways that the  first four
lightest sparticles may stack up in their mass hierarchy, 
one finds three such hierarchical mass patterns 
where the gluino is the NLSP  \cite{Feldman:2007fq,Feldman:2008hs} which have
been classified as models  (NUSP13, NUSP14, NUSP15) 
\footnote{There is another sparticle mass pattern NUSP10 \cite{Feldman:2008hs} 
with the mass hierarchy  $\na$ $<$ $\ta$ $<$ $\g$ $<$ $\cha$, 
 where  $\ta$ is the NLSP, but the $\tilde g$ lies close
to the $\ta$ mass.} 
as given in Table(\ref{gnlsptable1}). 
 We will often refer to this subclass of NUSUGRA as the GNLSP class of models.
Although progress has been made on the parameter space of sparticle masses 
with coannhilating gluinos \cite{Profumo,Bernal,Feldman:2007fq,Feldman:2008hs,bartol}, 
collider and dark matter detection implications of  the GNLSP models have yet to be explored in any great detail.
Thus in this work we give a dedicated analysis of such a model. 
 Since the gluino is a strongly 
interacting particle, an NLSP gluino will change drastically the typical sparticle analyses.
The phenomenology of GNLSP models is very different from that of a model where
the gluino is the LSP \cite{Baer:1998pg}
which we do not discuss in this paper. 
We note also that while relatively light gluinos have been studied in detail 
in reduced $SU(3)$ gaugino mass models \cite{Baer:2006dz},
the GNLSP situation, which we cover here, was not explored.
\begin{table}[h]
    \begin{center}
\begin{tabular}{c| lll}
 NUSP &   Mass Pattern  \\\hline
NUSP13  &   $\na$   $<$ $\g$    $<$ $\cha$ $\lesssim$ $\nb$     \cr
NUSP14  &   $\na$   $<$ $\g$    $<$ $\ta$ $<$ $\cha$         \cr
NUSP15  &   $\na$   $<$ $\g$    $<$ $A\sim H$             \cr

\end{tabular}
\caption{  Hierarchical sparticle  mass patterns for the four lightest sparticles, where $\tilde\chi^0 \equiv \tilde \chi_1^0$ is the LSP neutralino,
and where the gluino is
the NLSP  that arises in the NUSUGRA models. The labeling of the mass patterns is as given in \cite{Feldman:2007fq,Feldman:2008hs}.}
 \label{gnlsptable1}
\end{center}
 \end{table}
 
The outline of the rest of the paper is as follows: In Sec.(\ref{gaugino}) we discuss the origin of nonuniversalities
in the gaugino masses in $SU(5), SO(10)$ and $E_6$ GUT models. Thus this means that the 
ratio of the $U(1)$, $SU(2)$ and $SU(3)$ 
gaugino masses at the grand unification scale $M_G$ are  not in the  ratio $1:1:1$.
 Here  we point out that while no $F$ term breaking with a 
single irreducible representation can generate a GNLSP model, it is possible to do so with a mixture of  
two (or more) such breakings. Specifically we consider a linear combination of breaking with a singlet and  a non-singlet $F$ term
and show that several models exist which  lead to a GNLSP model. We also show  that there exists a subclass of models 
which superficially look different but are in fact isomorphic. In Sec.(\ref{experimental}) we discuss the techniques for
the computation of the sparticle spectrum at the weak scale and also discuss the  experimental constraints that are imposed
on the spectrum.  
 In Sec.(\ref{relic}) we give a discussion of how the relic density consistent with WMAP data is satisfied under the assumption that dark matter is entirely constituted of cold dark matter in
 the form of R parity odd LSP neutralinos. 
Here  it is shown that there are two main mechanisms by which this can come about. 
The first  mechanism is by coannihilation
 with gluinos where the dominant processes which participate in the coannihilation are $\tilde \chi^0\tilde \chi^0 \to f\bar f$,
 $\tilde \chi^0 \tilde g\to q\bar q, ~\tilde g \tilde g\to gg, q\bar q$. The second mechanism is the one where 
the LSP has a significant higgsino component and here the relic density constraint is satisfied in a
similar fashion as in the usual higgsino dominated LSP model.
However, this is rather rare, and when it occurs, it is often with a small amount of gluino coannihilation.
In Sec.(\ref{parameter}) we delineate the allowed parameter space for the GNLSP models and show that  there is 
a significant region of the parameter space where such  models manifest. 
In Sec.(\ref{compression}) we show that the GNLSP models lead to a compressed sfermion spectrum for the first two 
generations. Specifically the sleptons and the squarks of the first two generations are essentially degenerate with
the sleptons sometimes being heavier than the squarks. 
In Sec.(\ref{signature}) we give an analysis of the signatures for  GNLSP models. Here we discuss three sets of 
post trigger level cuts labeled C1, C2 and C3 
which are designed to reduce the background and enhance the signal to background ratio.
It is found that the dominant signatures are jets and missing energy and with properly chosen post trigger level 
cuts they stand out above the background. It is further found that a discovery of a GNLSP model can come about 
with an integrated luminosity of 10 fb$^{-1}$ at the LHC with a gluino mass of up to 800 GeV. More generally,
the models discussed here can be put 
to test with the first data from the LHC.

   In Sec.(\ref{dark}) we discuss the direct detection of dark matter in  GNLSP models. 
  It is found that the CDMS-08 data already constrains the parameter space of  GNLSP models,
  although only rather mildly  at the level of $\sigma_{SI}(\tilde\chi^0 p)\simeq 10^{-44}$cm$^{2}$. 
    Further, the future data from CDMS and LUX will either detect dark matter predicted in 
  this model or constrain the parameter space of the model. It is also noted, however, that a part of the
  parameter space of the model leads to a rather small spin independent neutralino-proton cross sections,  i.e.,
  $\sigma_{SI}({\tilde\chi^0 p})<10^{-46}{\rm cm}^{2}$, which lies outside the reach of the current direct detection 
   experiments and similar experiments in the foreseeable future. Interestingly, much of this parameter space will be accessible
  at the LHC since the gluinos can be produced 
 and detected via their jet and missing energy signatures
  as discussed in Sec.(\ref{signature}). In this sense, the LHC and the  direct detection experiments are complementary.
 In Sec.(\ref{benchmarks})
  we discuss the benchmarks for the three GNLSP model sets A, B and C.   

   Conclusions are given  in Sec.(\ref{conclusions}).
    In  Appendix A we give sum rules on the gaugino masses that hold for the various cases of 
   nonuniversalities that appear in Table (\ref{nonunitable1}). These sum rules also hold when one includes
   a singlet breaking along with breaking with a non-singlet. In Appendix B we give an analysis 
   at one loop which explains the compression of the sfermion spectrum for the first
   two generations. Benchmarks discussed in Sec.(\ref{benchmarks}) are given in Appendix C.
    
\section{Gaugino mass nonuniversalities in GUT models, the gluino NLSP 
and scaling \label{gaugino}} 
\.4
There is considerable literature on nonuniversalities of soft breaking and their
applications\cite{NU} within the framework of supergravity grand 
unification\cite{msugra,hlw}.
Our focus will be on the gluino phenomenology that results from the gluino being an NLSP
[For recent  analyses related to  gluino phenomenology in various contexts see
\cite{Arvanitaki:2005fa}]. 
Specifically, our focus here will be on nonuniversalities in the 
  gaugino mass sector arising from  $F$ type breaking in $SU(5), SO(10)$, and $E_6$ GUT groups
which have been discussed over the years\cite{Ellis:1985jn,nonuni2} and a more comprehensive analysis has been given recently\cite{Martin:2009ad}. 
 Results of this analysis 
 are  summarized in Table(\ref{nonunitable1})
 \footnote{In this analysis we do not consider the flipped
  models and the ratios  listed in 
  Table (\ref{nonunitable1}) exclude such models.}.
  In the table, ratios of gaugino masses that arise  when the GUT symmetry is broken by
an $F$ term, which is an irreducible representation of the gauge group $SU(5)$, $SO(10)$,
and $E_6$, and enters in the decomposition of the symmetric product of two 
adjoint representations corresponding to the relevant group. Table(\ref{nonunitable1}) identifies
the group and the irreducible representation and the corresponding ratio of the gaugino masses. 
For $SO(10)$ and $E_6$ several gaugino mass ratios are listed for a given irreducible representation. 
These correspond to different patterns by which the GUT symmetry breaks to lower rank groups.
Further details can be found in \cite{Martin:2009ad}.
   \begin{table}[h]
    \begin{center}
\begin{tabular}{|c| ccc|c|ccc|    }
\hline
Group       &   Rep.            &   Label    &    $M_1:M_2:M_3$         & Group  &   Rep.         &     Label     &    $M_1:M_2:M_3$ \\\hline
$SU(5)$   	&	 {\bf 1}      	&	-	&	 $1:1:1$               & $E_6$ &	 {\bf 650}   &	   (12)	&	 $-1:1:1$      \cr                   	
          	&	 {\bf 24}     	&	(1)	&	 $-1/2$:$-{3}/{2}:1$   &	    &	             &	 (13)	&	 $-1:1:0$    \cr                  	
          	&	 {\bf 75}     	&	(2)	&	 $-5:3:1$              &        &	        	 &		(14)	&	 ${1}/{10}:-{3}/{2}:1$ \cr 
          	&	 {\bf  200}   	&	(3)	&	 $10:2:1$              &	    &	             &		(15)	&	 $-{13}/{5}: 1:1$ \cr	
$SO(10)$  	&	 {\bf 210}    	&	(4)	&	 $-3/5:1:0$            &     	&	        	 &	(16)	&	  ${1}/{5}:1:0$ \cr  	
          	&	              	&	(5)	&	 $-4/5:0:1$            &      	&	        	 &	(17)	&	 ${41}/{15}:1:1$ \cr
          	&	              	&	(6)	&	 $1:0:0$               &  &	 {\bf 2430}  &  	(18)	&	 $-{11}/{5}:1:0$ \cr	
          	&	 {\bf 770}    	&	(7)	&	 $19/{10}:{5}/{2}:1$   &	    &	             &	(19)	&$1:{35}/{9}:1$ 	 \cr 
          	&	              	&	(8)	&	 $ {32}/{5}:0:0$       &	    &	             &	 	(20)	&${12}/{5}:0:1$ 	  \cr
$E_6$    	&	 {\bf 650}    	&	(9)&	 $-{1}/{5}:1:0$        &	    &	             &	(21)	& $0:0:1$  	 \cr		
          	&	              	&	(10)&	 $-{1}/{5}:-1:1$       &	    &	             &	 	(22)	&	${33}/{5}:1:1$ \cr
          	&	              	&	(11)	&	$3:1:1$	                   &	    &	      &  (23)	& $9/{5}:1:0$\cr    
\hline	
\end{tabular}
\caption{Exhibition of the gaugino mass ratios at the GUT scale for various groups and representations in
$SU(5), SO(10)$, and $E_6$ models\cite{Martin:2009ad}.
The mass ratios are  listed in a hierarchical manner, i.e.,  they are listed in the order of the 
smallest rank group and lowest dimensional representation in which they first appear and are labeled
from (1)-(23).
Thus a specific ratio may be repeated several times as one goes  up the chain. }
 \label{nonunitable1}
\end{center}
 \end{table}
 None of the models listed in Table(\ref{nonunitable1}) can give rise to a
  gluino as the  NLSP with $F$ type breaking with a single irreducible  representation\footnote{This also holds for 
  flipped models, i.e.,  $F$ type breaking with a single irreducible representation cannot give rise to a gluino as the NLSP.}. However, we will show that a combination of GUT symmetry breaking
in the gaugino mass term sector 
with two irreducible representations does allow for a gluino as the NLSP for a subset of models
  listed in Table(\ref{nonunitable1}). Specifically we will consider a linear combination of 
  a singlet and a non-singlet $F$ term. In this case an  interesting phenomenon arises in that the models with the same value 
$r\equiv (M_2-M_1)/(M_3-M_1)$
  can be made isomorphic under redefinitions and scalings in the gaugino sector. 
Thus suppose we write the gaugino masses for models  of the above  type with a singlet and
a non-singlet F breaking so that
  \beqn
   M_1^{(i)}= (1+a_i \alpha_i)\mh,
     ~M_2^{(i)}= (1+b_i \alpha_i)\mh,
       ~M_3^{(i)}= (1+c_i \alpha_i)\mh,
       \label{non1}
  \eeqn
  where the first term within each of the parentheses on the right hand side in Eq.(\ref{non1}) arises
  from the singlet contribution, and the second term within each of the parentheses is  the contribution
  from the non-singlet. Here 
   $i$ defines a specific model and 
$a_i, b_i, c_i$ are the fractions given in Table(\ref{nonunitable1}) with $\alpha_i$ being an
 arbitrary parameter. 
 Next we note that two models $i$ and $j$ defined by Eq.(\ref{non1}) 
 can be made isomorphic if they have the same value of $r$ in the sense that 
  \beqn
M_{a}^{(i)} = \lambda_{ij} M_{a}^{(j)}; a=1,2,3,
\eeqn 
when $\alpha_j$ is related to $\alpha_i$ in the  following way
\beqn
\alpha_j^{-1}({b_i-a_i})=  \alpha_i^{-1}(b_j-a_j) +  a_ib_j-b_ia_j.
\label{ab}
\eeqn
 This means that under the constraint of Eq.(\ref{ab}), a rescaling of $\mh$ of model $j$ can make it
 isomorphic to model $i$. Thus in essence,  models
 with the same value of 
 $r$ would in fact be equivalent when taken  in a linear combination of breakings including singlets. 
Using Eq.(\ref{non1}) and Table(\ref{nonunitable1}) one finds that there are 
several possibilities for which the GNLSP class of models can arise. We limit ourselves to the
following cases:
   \begin{enumerate}
\item
 Model \a (ISO-I) : This class of models arise where $r$ takes the common value $-2/3$ as exhibited below
\begin{equation}
\left. \begin{array}{c} M_1:M_2:M_3 \\ \hline  -1/2:-3/2:1 \\ 19/10:5/2:1 \\ 
-1/5:-1:1  \end{array} \right\}\longrightarrow \begin{array}{c} r=-2/3.
\end{array}
\label{iso1}
 \end{equation}

\item
 Model \b: This is an $E_6$ model with $F$ type breaking with 
 {\bf 2430} plet such that\cite{Martin:2009ad}  $E_6\to SU(6)''\times SU(2)_L( 2430\to (189,1))$
 which gives  $M_1:M_2:M_3=0:0:1$. This model can generate a gluino 
 as the NLSP upon the addition of breaking with a singlet 
\footnote{We note that there is another $E_6$ model 
 with $F$ type breaking with
  {\bf 2430} plet such that $E_6\to SU(6)''\times SU(2)_L( 2430\to (405,1))$\cite{Martin:2009ad}
 which gives  $M_1:M_2:M_3=\frac{12}{5}:0:1$ ($r=12/7$). 
 This model can also generate a gluino 
 as the NLSP upon  addition of a singlet and there is 
 a relative sign flip between $M_1$ and $M_2$ in this case.  
  However, the model gives a light Higgs mass in the parameter space investigated
  which falls below the 
 current limits and thus we do not consider this model further.}.
 \item
Model \c: Here $r$ is free and thus defining $r=\delta_2/\delta_3 $ the gaugino masses
at the GUT scale may be parametrized as
\be
 {\widetilde M}_1 = \mh, ~~~{\widetilde M}_2 = (1+\delta_2)\mh,~~~{\widetilde M}_3 = (1+\delta_3)\mh,
\ee
and  $\delta_2$ and $\delta_3$
 can be varied independently. Model \c contains models GNLSP$_A$ and \b as subcases\footnote{We remark that in \cite{Feldman:2007fq,Feldman:2008hs}, where the gluino NLSP
in SUGRA models was previously observed,
the notation $\delta_5,\delta_6$ was used. }. 
\end{enumerate}   
Aside from the model discussed in footnote 4, models \a, \b, and \c are  
 the only  models  which lead to a GNLSP through breaking with a singlet and a nonsinglet.
This can be seen easily by using the semi analytic analysis
given in Appendix B.
For all the three models a GNLSP  requires $\delta_3$ to lie in the range $(-0.9,-0.8)$. 
Some benchmarks for Models A,B and C are given in Tables(\ref{benchAin},\ref{benchBin},\ref{benchGin}) and a display of their partial sparticle  spectrum and  some other properties of these models are exhibited in  Tables(\ref{benchAout},\ref{benchBout},\ref{benchGout}). 
We also note that from  the analysis of \cite{Martin:2009ad} one can discern another set of models which have 
 the same common value of $r$. Thus the models 
 with the gaugino mass ratios
$M_1:M_2:M_3$=  (i)$-\frac{1}{5}:3:1$;  (ii)$\frac{2}{5}:2:1$;
(iii)$-\frac{3}{5}:1:0$;(iv) $\frac{5}{2}:-\frac{3}{2}:1$;
(v) $\frac{1}{10}:\frac{5}{2}:1$;(vi) $\frac{8}{5}:0:1$ have the 
common value $r=8/3$. One  may call this ISO-II because when combined with a singlet $F$ type breaking these models
too would be isomorphic so that the six different models are effectively one model
as far as the gaugino sector is concerned. However, this model class does not lead to
a GNLSP  which is the main focus of this paper.
In the  following we discuss  the GNLSP models in further detail including the satisfaction of the
relic density, the production cross section of the gluinos,   the
 signatures for their identification at the LHC, and the direct detection of dark matter in the GNLSP class of models.
\.4
\section{Experimental constraints\label{experimental}}
\.4
\label{exp}
 Our general 
 procedure is similar to that  discussed in \cite{Feldman:2008hs} which we briefly describe 
 below. 
 In the analysis one specifies boundary conditions of the model 
 at the grand unification (GUT) 
 scale which we take to be $M_{G}\sim 2\times 10^{16}$ GeV.
 Specifically
   we take the sfermion masses at the GUT scale to be universal, but assume that
 the gaugino masses are in general nonuniversal with nonuniversalities given by 
 $\delta_2$ and $\delta_3$. One then uses renormalization group equations (RGEs)
to compute the sparticle mass matrices and their eigenvalues at the electroweak scale.
The code  used in these RGE evolutions and computations of the sparticle spectrum is  SuSpect2.41
\cite{SUSPECT},  and similar results are obtained with SoftSUSY\cite{ben} and SPheno \cite{SPHENO}. 
Further, one imposes the lower limit constraints on the sparticle masses from the LEP and from the Tevatron data 
as well as constraints from the WMAP on the relic 
density.  The analysis of the relic density is first done at the perturbative level  with 
MicrOMEGAs \cite{MICRO}, which relies on CalcHEP \cite{Pukhov:2004ca}.   Non-perturbative effects on the relic
density are also discussed.

Below we give a list of the relevant constraints from collider and astrophysical data which
have been included in the analysis:

(i)  The 5-year WMAP data  constrains the 
relic density of dark matter in the universe so that  $\Omega_{\rm DM} h^2 = 0.1131\pm 0.0034$ \cite{Komatsu:2008hk}.
We take a 6$\sigma$ corridor around the central value to constrain the 
relic abundance of neutralinos.
The larger band
is taken due to the sensitivity of the relic density computation in particular regions 
of the parameter space. A large class of our models fall well within a $2\sigma$ bound. 

(ii)
The FCNC process $b\rightarrow s\gamma$ receives a significant contribution 
from the SUSY processes\cite{susybsgamma}. 
The Heavy Flavor Averaging Group (HFAG) \cite{Barberio:2008fa} along with the
BABAR, Belle and CLEO give experimental results: 
${\mathcal Br}(B \to X_s \gamma) =(352\pm 23\pm 9) \times 10^{-6}$.
A new estimate of standard model contributions at
$O(\alpha^2_s)$ gives \cite{Misiak:2006zs} ${\mathcal
Br}(b\rightarrow s\gamma) =(3.15\pm 0.23) \times 10^{-4}$. We utilize both 
experimental and theoretical progress in the evaluation of this observable and take a 3$\sigma$ 
corridor around the experimental value,  $2.77\times 10^{-4}<{\mathcal
Br}(b\rightarrow s\gamma) <4.27\times 10^{-4}$, to constrain the theoretical prediction 
including both SM and SUSY contributions. 

(iii)
Another important constraint from B-physics is the rare decay process 
$B_s\to \mu^+\mu^-$ which can become significant for large $\tan\beta$\cite{bsmumu}. 
The most stringent 95\% (90\%) C.L. limits are achieved by CDF\cite{2007kv}
${\mathcal Br}( B_s \to \mu^{+}\mu^{-})<$ $5.8 \times 10^{-8}$ ($4.7 \times 10^{-8}$).
We take a conservative limit ${\mathcal Br}( B_s \to \mu^{+}\mu^{-})  < 10^{-7}$.

(iv) For the constraints from the anomalous magnetic moment of the muon, 
we use a conservative bound $-11.4\times 10^{-10}< \delta(g_{\mu}-2)<9.4\times 10^{-9}$ 
as in \cite{Djouadi:2006be} where $\delta(g_{\mu}-2)$ is the new physics 
contribution to $(g_{\mu}-2)$ beyond the standard model.

 (v)
 Additionally, we also impose various mass limits as follows:
 $m_{\cha}>104.5 ~{\rm GeV}$ 
\cite{lepcharg} for the lighter chargino,  $m_{\ta}>101.5 ~{\rm GeV}$  
for the lighter stop,  and $m_{\sta}>98.8 ~{\rm  GeV}$ for the lighter stau. 
For the lightest  CP even  Higgs boson mass in MSSM 
we take the constraint to be $m_h> 100 ~{\rm GeV}$ (90\% of the models
that pass all constraints have  $m_h> 110 ~{\rm GeV}$). One may compare these with the
 standard model like Higgs boson mass limit which is 
$\approx$ 114.4 {~\rm GeV} \cite{smhiggs}. 
 For the gluino mass, recent Tevatron experiments give 
$m_{\g}>308 ~{\rm  GeV}$ (D-Zero)\cite{:2007ww} 
and $m_{\g}>280 ~{\rm  GeV}$ (CDF) \cite{Aaltonen:2008rv}.
The limits given by  \cite{:2007ww,Aaltonen:2008rv} are valid 
within the framework of the minimal supergravity models and may be modified
in nonuniversal SUGRA models. 
Hence the total SUSY 
production cross section constrained by the Tevatron analyses 
will typically be a larger total cross section than that which arises in the GNLSP models.
Further, as we will show shortly, the mass splitting between the NLSP gluino and LSP neutralino
must be relatively small in order to satisfy relic density constraints. Thus the 
relatively small mass splittings between the LSP and GNSLP 
can lead to softer decay products and an overall lower multiplicity of final state events relative
to models for which the mass splitting is significantly larger.
Therefore in this analysis,  we take a conservative  lower bound, namely
$m_{\g}>220 ~{\rm  GeV}$. 
Our choice of this lower bound is taken as to not eliminate a part of the parameter space which may otherwise
be allowed pending a full analysis of the Tevatron data using nonuniversalities (see also \cite{Belyaev:2007xa} for a related discussion regarding a lower
bound on the mass of the gluino). 
\.4
 \section{Relic density via gluino coannihilation\label{relic}}
 \.4
It is interesting to ask how the relic density constraints are satisfied
in the class of models  with the gluino as the NLSP as these constraints have important implications for collider phenomenology (for recent works  connecting  sparticle phenomenology at colliders and
 dark matter see \cite{arno,Feldman:2008en,Bhattacharyya:2008zi,Feldman:2008jy,Biswas:2009zp}). As an illustration we consider the model
\be 
{\rm GNLSP}_{\rm Co}: (m_0, \mh, A_0, \tan \beta,\delta_2,\delta_3) = (1450, 730, 2700, 40, 0.332, -0.839 ), 
\label{co}
\ee
where all masses are in GeV and $sign(\mu)$ is taken to be positive. We will take the top mass at 170.9 GeV throughout this work, though the analysis
here does not show great sensitivity to the top mass. 
This model gives $(m_{\na},m_{\g})= (305.1,348.6)$ GeV.
For $\rm GNLSP_{\rm Co}$ 
the channels which contribute to $1/(\Omega h^2)_{\tilde\chi^0}$  more than $1\%$ are as follows :
 $  \g \g \to g g (47\%)$,
 $  \g \g \to u \bar u (8\%)$,
 $  \g \g \to c\bar c (8\%)$,
 $ \g \g \to d\bar d  (8\%)$,
 $  \g \g \to  s \bar s (8\%)$,
 $ \g \g \to b  \bar b (6\%) $,
 $  \g \g \to t \bar t (4\%)$,
  $\na \na \to  b  \bar b (6\%) $,
  $ \na \g \to t \bar t (2\%) $,
  $ \na \na \to t \bar t (2\%)$,
 $ \na \na \to \tau^{+}  \tau^{-}  (1\%)$.
The relic density is $(\Omega h^2)_{\tilde\chi^0} = 0.108$ at the 
perturbative level, and the model
has eigen decomposition $\na = 0.986{\tilde b} -0.016{\tilde w} +0.146{\tilde h}_1 -0.092 {\tilde h}_2$ where $\tilde b, \tilde w$,  are the bino and wino components and $\tilde h_1, \tilde h_2$ are the higgsino components, and thus the model has a substantial higgsino component. 
This model belongs to the pattern classified as  NUSP13 in\cite{Feldman:2007fq,Feldman:2008hs}.

From the above it is clear that the gluino processes dominate the WIMP annihilation 
at the freezeout temperature in the early universe\cite{Profumo}.
Further the LSP mass and the NLSP mass are close with  a mass difference
$\Delta_{\tilde g\tilde \chi^0}\equiv(m_{\tilde g}- m_{\na})/m_{\na}\approx 0.14$.
An examination of the mass 
splittings and the associated annihilation processes point to 
 a strong coannihilation occurring in the model of Eq.(\ref{co}).
 Thus consider the annihilation processes $\tilde \chi_i\tilde \chi_j$ going into the standard 
 model particles. Here 
 the effects of 
 coannihilation are controlled by the  Boltzmann
suppression factor\cite{Griest:1990kh}
 \beqn \gamma_i=\frac{n_i^{\rm eq}}{n^{\rm eq}} =
\frac{g_i(1+\Delta_i)^{3/2} e^{-\Delta_i x}} {\sum_j g_j
(1+\Delta_j)^{3/2}e^{-\Delta_j x}}\ , 
\label{coann}
\eeqn 
where $g_i$ are the
degrees of freedom of $\chi_i$, $x={m_1}/{T}$
and $\Delta_i =(m_i-m_1)/m_1$, with $m_1$ defined as the 
LSP mass.  
The processes which 
dominate the WIMP annihilation in the early universe are 
\beqn
\na\na \to F,
~\na\tilde g \to F', 
~\tilde g \tilde g \to F''~,
\eeqn
where $F, F', F''$ constitute the pairs of standard model states. 
The relic density is controlled by the integral  
\beqn
J_{x_f}=\int_{x_f}^{\infty} x^{-2} \langle \sigma_{\rm eff} v \rangle dx~,
\label{relic1}
\eeqn
where $v$ is the relative velocity of annihilating  supersymmetric particles,
$\langle \sigma_{\rm eff}v \rangle$ is the thermally averaged cross section times the relative velocity and $x_f$ is the freezeout temperature.
The 
$\sigma_{\rm eff}$ that enters the relic density can be written approximately as follows 
\beqn
\sigma_{\rm eff}\simeq \sigma_{\tilde g \tilde g} \gamma^2_{\na}\left(\gamma^2 + 2\gamma \frac{\sigma_{\na\tilde g}}{\sigma_{\tilde g \tilde g}} + 
\frac{\sigma_{\na\na}}{\sigma_{\tilde g \tilde g} }\right)~,
\label{relic2}
\eeqn
where $\gamma =\gamma_{\tilde g}/ \gamma_{\na}$ and where $\gamma_i$ are defined by Eq.(\ref{coann})
and where \cite{Baer:1998pg}
\beqn
\sigma(\g\g \to gg)&=&{3\pi\alpha_s^2\over 16\beta^2s}\left\{\log{1+\beta\over
1-\beta}\left[21-6\beta^2-3\beta^4\right]-33\beta+17\beta^3\right\},\nonumber\\
\sigma(\g\g\to q\bar q)&=&{\pi\alpha_s^2\bar\beta\over 16\beta s}
(3-\beta^2)(3-\bar\beta^2)\,.
\label{xsecgg}
\eeqn
Here $\beta = \sqrt{1-4 m^2_{\g}/s}$,  and
 the quark mass enters Eq.(\ref{xsecgg}) 
 through $\bar\beta=\sqrt{1-4m_q^2/s}$.
One interesting phenomenon concerns the following: we know that the cross section for the annihilating 
gluinos falls  with the gluino mass. 
On the other hand $\sigma_{\rm eff}$ that enters the relic 
density analysis must be nearly constant for a  wide range of gluino masses so that the relic density
be satisfied. This can happen due to the presence of the  
coannihilation factor $\gamma^2$ which multiplies 
$\sigma_{\tilde g\tilde g}$ in Eq.(\ref{relic2}). 
This is easily seen by noticing that 
 $\Delta_{\tilde g\tilde \chi^0}=(m_{\tilde g}-m_{\tilde\chi^0})/m_{\tilde\chi^0}$
has a dependence on the gluino mass of the form \footnote{A similar relationship in a graphical form appears in the
analysis of \cite{Profumo} for a bino LSP.}
\beqn
\Delta_{\tilde g\tilde\chi^0}^0 -C ~log(m_{\tilde g}/m_{\tilde g}^0)~,
\label{relic3}
\eeqn
where $\Delta_{\tilde g\tilde\chi^0}^0= (m_{\tilde g}^0-m_{\tilde\chi^0}^0)/m_{\tilde\chi^0}^0$,
 and $m_{\tilde g}^0$ is a reference 
gluino mass and $m_{\tilde \chi^0}^0$ is the corresponding reference neutralino mass, and $C>0$.
What one finds is that the difference $\Delta_{\tilde g\tilde\chi^0}$ decreases when gluino mass increases
which enhances $\gamma$ and compensates for the falling cross section $\sigma_{\tilde  g\tilde g}$.
The above phenomenon sustains an essentially constant $J_{x_f}$ as the gluino mass varies
allowing for a satisfaction of the relic density over  a wide range of gluino masses. 
 What the analysis implies is that
 $m_{\tilde g}/m_{\tilde \chi^0}$   tends to unity 
as the gluino mass increases.  A numerical analysis
bears this out. 
It is, however, interesting to note, that we also find  few
cases where the GNLSP emerges without significant 
\co which occurs when the LSP 
has a significant higgsino component which allows for the satisfaction
of the relic density constraint in a manner quite similar to what happens on the
Hyperbolic Branch of REWSB\cite{hb}. Indeed, there are cases where 
the neutralino annihilations are seen to dominate the annihilation cross
sections via $\na \na \to (b \bar b, \tau^{+} \tau^{-})$ with only a small contribution to 
the satisfaction of the relic density constraints
arising from LSP-GNLSP coannihilation, and in some cases coannihilation enters only at the single percent level
 (an example is ${\rm GNLSP}_{\rm C1}$ given in Appendix C). Another interesting example is 
 model ${\rm GNLSP}_{\rm A1}$ (also given in Appendix C) which proceeds with annihilation contributions to relic density calculation 
dominantly via 
(62\%)$ \na \na \to t\bar t$, and  (15\%) $\na \na \to W^{+}W^{-}$, 
and only a small fraction 
(3\%) for $\g \g \to g g $ and the remainder coming from neutral diboson final states.
\.4

We discuss now possible  
nonperturbative corrections to the 
annihilation cross section.
As shown in Refs \cite{Baer:1998pg,Profumo} 
nonperturbative effects 
on the  annihilation cross section can be relevant near threshold
where 
multiple gluon exchange, for example, can give rise to the so-called
Sommerfeld enhancement factor $\cal E$.  These effects 
may be approximated as \cite{Baer:1998pg}
\beqn
{\cal E}_{j}={C_{j} \pi \alpha_s\over \beta}
\left[1-\exp\left\{-{C_{j} \pi \alpha_s\over \beta}\right\}\right]^{-1}\,,
\eeqn
where $C_{j=g}=1/2$ ($C_{j=q}=3/2$) 
for $\g \g \to gg$ ($\g\g \to q\bar q$) respectively, and we note that ${\cal E}_{j}$
enters bilinearly in Eq.(\ref{xsecgg}).

Bound states can form as well if the gluino is a stable LSP. 
We do not consider the latter situation. However,  as already discussed, 
for the GNLSP at the 
perturbative level, 
the dominant contribution in the 
annihilation cross
section for most models arises from the 
gluino-gluino annihilation modes, and this occurs for 
$\Delta_{\tilde g\tilde \chi^0}\lesssim 0.2$.
Since micrOMEGAs performs the relic density analysis using only  
perturbative cross section, we have carried out an independent analysis of the relic density 
to include the effects of the Sommerfeld enhancement.  Our analysis gives 
results which are in agreement with the analysis of third reference of \cite{Profumo}.
We note here that an increase in $\Delta_{\g \na}$ in the range of $(2-3) \%$ is needed
 when the Sommerfeld enhancement of cross section is taken into account. 
Equivalently the effect of the  Sommerfeld enhancement can be recast
as a shift in the gluino mass for a fixed LSP mass
 in order to have the same relic density as 
for the 
perturbative case. Specifically, an upward shift of the gluino mass by a few GeV is needed. 
Thus for example, for the model $\rm GNLSP_{\rm Co}$  discussed above, the relic density
constraint is satisfied with the inclusion of  
non-perturbative effects with a 3 GeV upward shift of the $SU(3)$ 
gaugino mass at the GUT scale leading to an increase in 
$\Delta_{\tilde g\tilde \chi^0} =14\%  \to 16 \%$.
More generally, we  
find that numerically for the non-perturbative case, for fixed LSP mass,  $m_{\g}$ 
needs to be increased by  (3 to 6) GeV to achieve the same 
 relic density as for the 
 perturbative case. The above holds for gluino 
 masses in the range up to about 1 TeV.
In Appendix C
 we give benchmarks including the effects of the
Sommerfeld enhancement.

 \section{Consistent parameter space of the gluino NLSP \label{parameter}}

\begin{figure*}[t]
 \begin{center}
  \includegraphics[width=14cm,height=10cm]{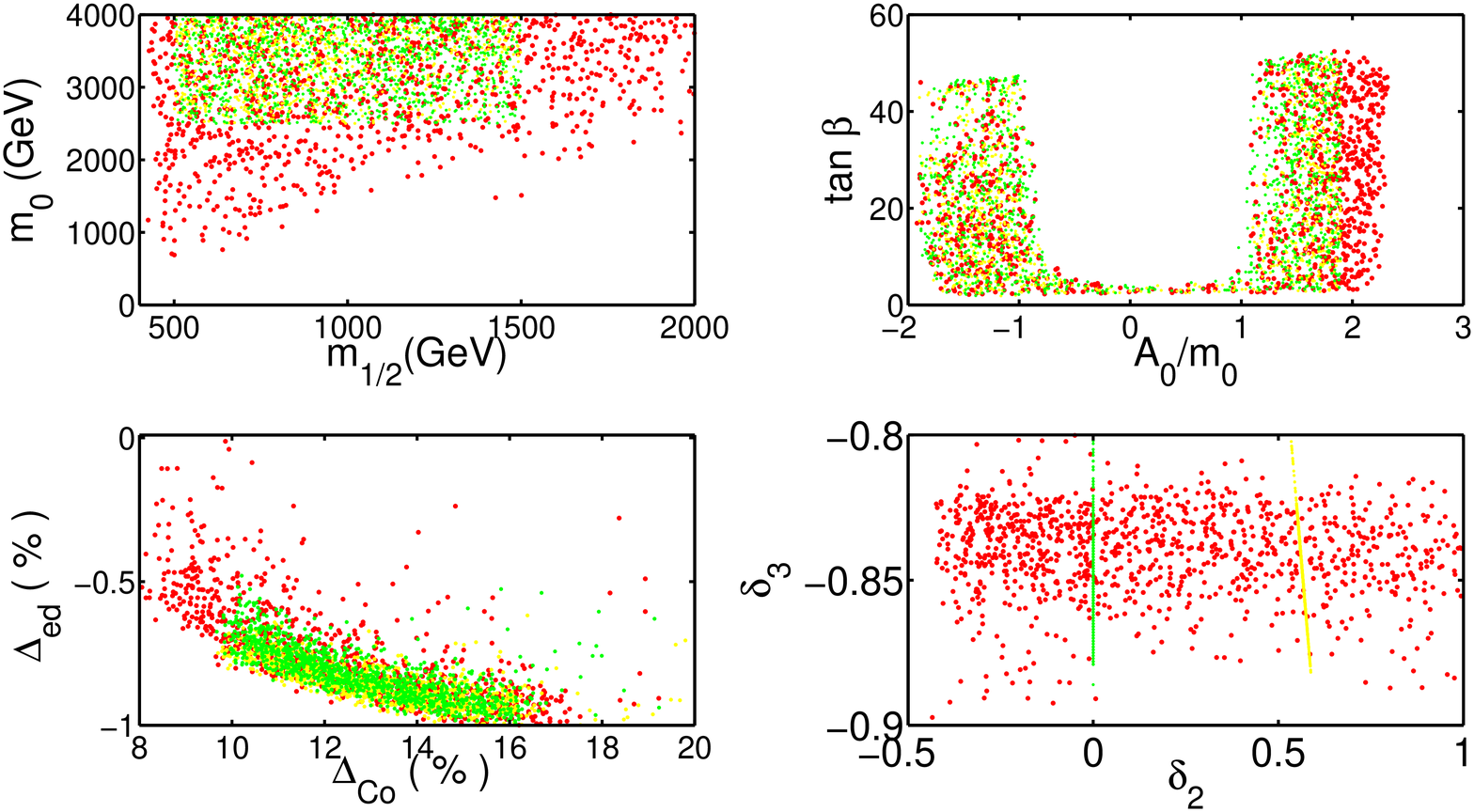}
\caption{(Color online) An analysis of the consistent parameter space in  GNLSP models.
Red (dark) model points are for Model \c, while  the yellow (light)  and the green
model points are for Model \a and Model \b as discussed in the text.
$\Delta_{ed}$ is the splitting of the slepton and squark masses in the first two generations
as also discussed in the text.
}
\label{a-tanbeta}
\end{center}
\end{figure*}
Based on the initial discovery of the existence of a viable parameter
space where the gluino is the NLSP in nonuniversal SUGRA models\cite{Feldman:2007fq,Feldman:2008hs}, we perform in this work  a dedicated search for delineating the parameter space of  GNLSP models 
consistent with the radiative electroweak symmetry breaking constraints and with the experimental
constraints from colliders and from the relic density.
In our analysis  the input parameters $m_0, \mh, A_0, \tan\beta,\delta_2,\delta_3$   assume the following bounds:
 $m_0 < 4$ TeV, $\mh < 2 $ TeV, $|A_0/m_0|<3$, and $\tan \beta \in (1,60)$,
while  
  $\delta_2$ and $\delta_3$  are
  chosen in a manner appropriate for models A, B and C defined in Sec.(\ref{gaugino}).
 Thus for model \a,  $\delta_2$  is determined by the constraint $r=-2/3$, for model \b,  $\delta_2 =0$ and
 for the  model \c, $\delta_2$  is assumed to lie in the range $\delta_2 \in (-0.9,1)$,
 while $\delta_3$ typically lies in the range $\delta_3\in (-0.9, -0.8)$.
 Within the ranges assumed above, we find a  significant region of the parameter space where 
 each model is  realized. Of course, the parameter space for model \c is larger than that 
 for the model \a or for the model \b, but the parameter space for models \a and \b are also quite significant as shown in  Fig.(\ref{a-tanbeta}).
  Some interesting observations can be made from the analysis of
Fig.(\ref{a-tanbeta}). Thus the top left  panel of Fig.(\ref{a-tanbeta}) shows that
typically $m_0>\mh$ for this class of models while the top
 right panel shows  that the region $A_0/m_0=0$ is very thinly populated which
is in sharp contrast to the mSUGRA case where the $A_0/m_0$ region is heavily populated.
The lower right panel of Fig.(\ref{a-tanbeta}) displays the allowed model points 
in the  $\delta_3$ vs $\delta_2$ plane which shows 
that  GNLSP models constrain the nonuniversality $\delta_3$ to lie in a very narrow range $(-0.9,-0.8)$\cite{Feldman:2007fq,Feldman:2008hs}
while $\delta_2$ is widely dispersed for model \c but restricted for models \a and \b
since $\delta_2/ \delta_3$ = $- 2/3$ for model \a and $\delta_2=0$ for model \b. 
These points are indicated in yellow (model GNLSP$_{\rm A}$) and green (model GNLSP$_{\rm B}$).
\begin{figure*}[t]
  \begin{center}
\includegraphics[width=8cm,height=6.5cm]{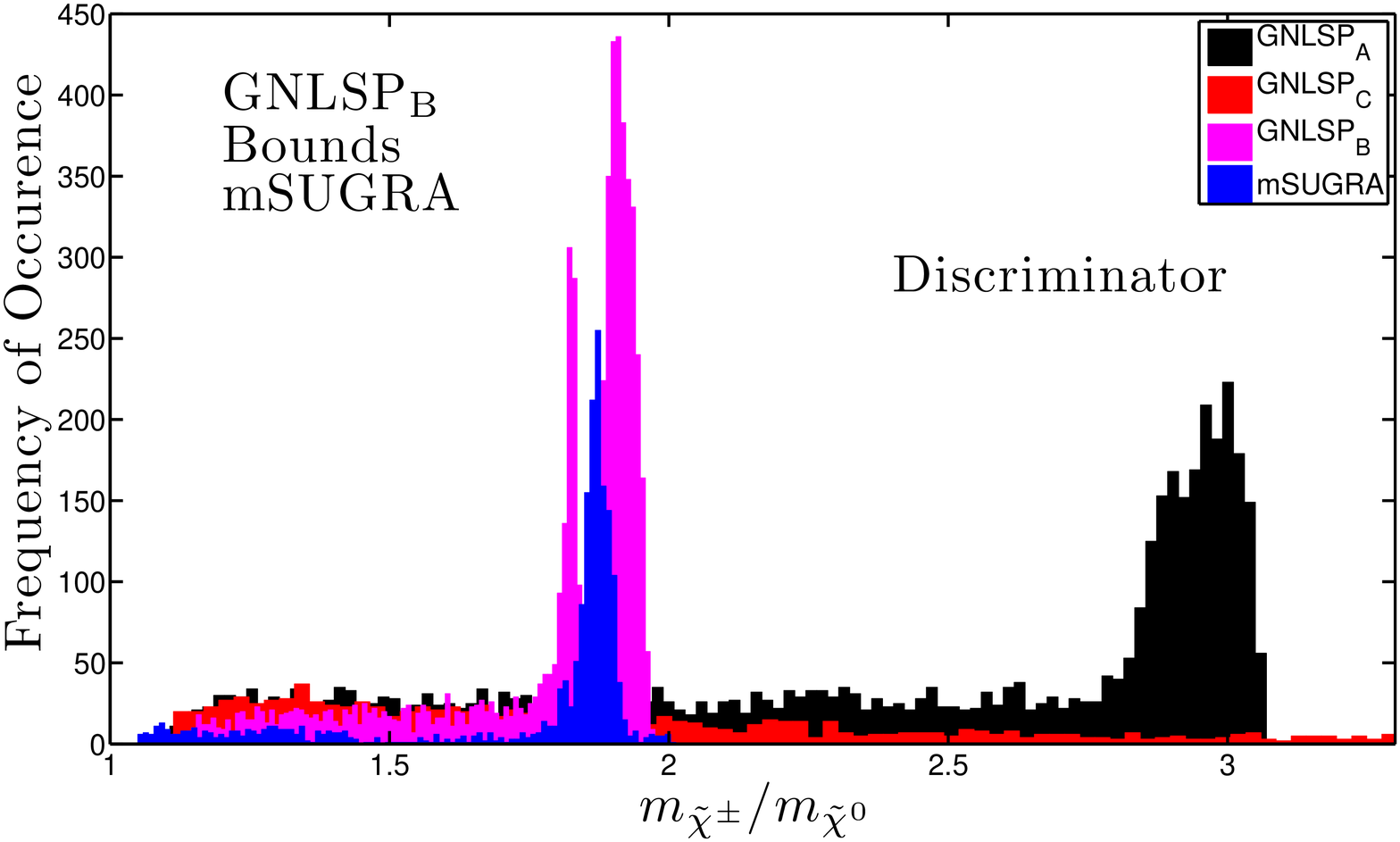}
\includegraphics[width=8cm,height=6.5cm]{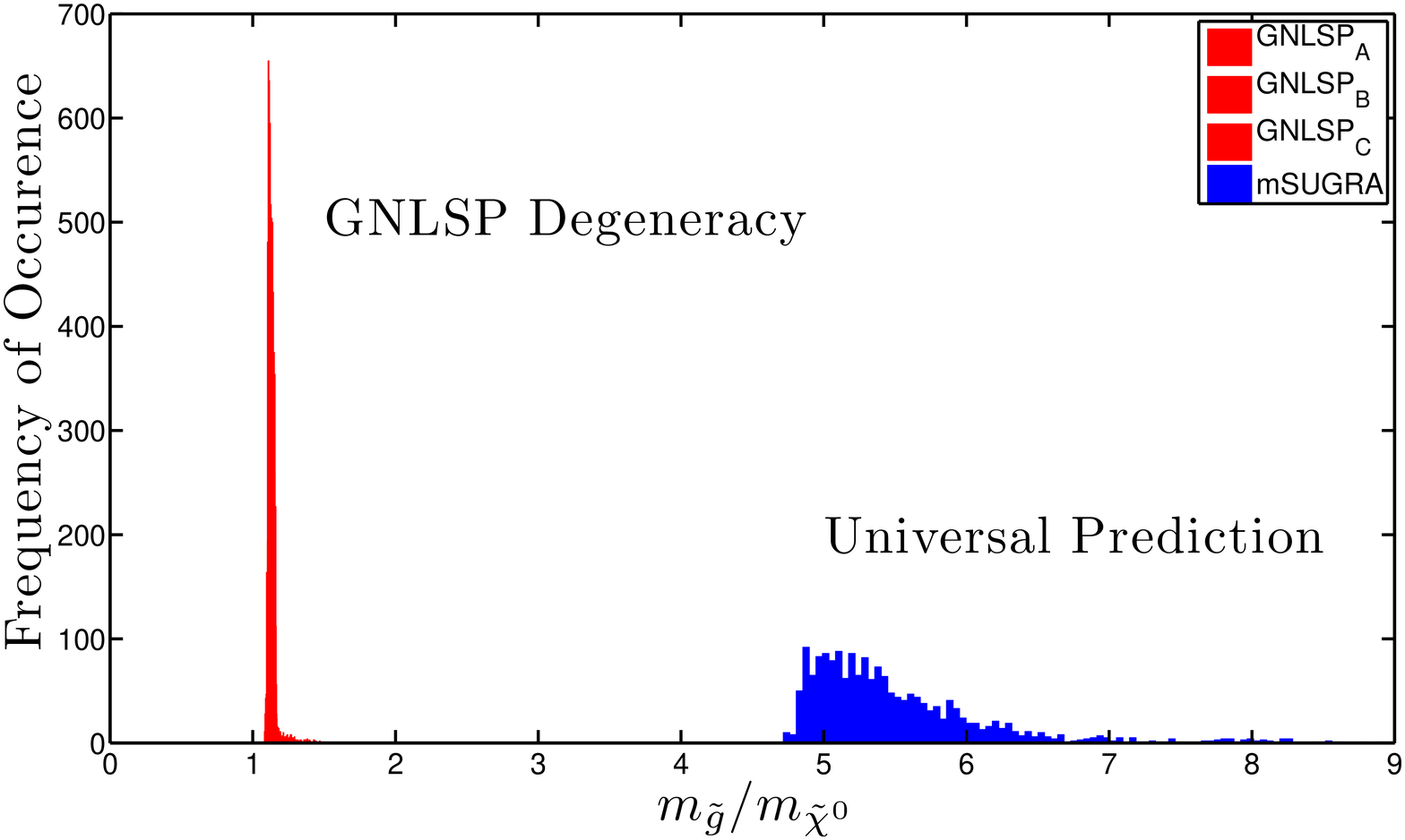}
\caption{(Color online) Left panel:
An exhibition of the scaling between the light chargino mass and the LSP mass for the GNLSP models 
\a, \b and \c vs the mSUGRA model. The figure shows that \a   produces
the ratio $m_{\tilde\chi^{\pm}}/m_{\tilde\chi^{0}}\sim 3$ which differentiates it from the bino branch
of mSUGRA and for GNLSP$_{\rm B}$.
Right panel: An exhibition of scaling in $m_{\tilde g}/m_{\tilde\chi^{0}}$. All GNLSP models 
are well separated from mSUGRA in this figure.}
 \label{scaling}
  \end{center}
\end{figure*}
\.4
\section{Compression of the sparticle mass spectrum in  GNLSP models\label{compression}}
\.4
In models with universal boundary conditions at the GUT scale for the gaugino masses, the 
gluino mass will be typically a factor of 5-6 larger than the lightest neutralino mass. A large
gluino mass tends to contribute a significant portion to the squark masses in the RG running. 
Thus in the mSUGRA model\cite{msugra} there is typically a significant splitting of the slepton and squark masses
for regions of the parameter space where $m_0$ and $m_{1/2}$ are comparable. 
This is generally not the case in the model under consideration where gluino is the NLSP. Here the 
gluino mass will be typically much lighter relative to the squark masses, and thus splittings between sleptons and squarks will 
be less pronounced. Specifically for the first two generations one will find  a rather compressed
spectrum. Table(\ref{sumrule}) exhibits the high  degree of  degeneracy of the squarks and of 
the sleptons in  the first two generations in models with the gluino as the NLSP relative to, e.g., 
the mSUGRA model [For a general discussion of sum rules see \cite{Martin:1993ft}].
 In the examples shown one finds that while for the mSUGRA 
 mSP3\footnote{The mSUGRA pattern mSP3 has the following hierarchy for the first four sparticles: $\tilde\chi^0<\tilde\chi_1^{\pm} <\tilde\chi_2^0<\tilde \tau_1$.
The largeness of the sfermion masses indicates that the electroweak symmetry breaking is realized on the
Hyperbolic Branch\cite{hb} (for recent works on the HB see \cite{Baer:2007ya,Feldman:2008jy}) .} 
  case 
the splitting between the sum of the down type quarks and the charged sleptons in the first
generation can be as much as 35\%, while for the GNLSP case  it is only about 
1\%. Further, while for the mSUGRA case the first and the second generation squarks are invariably
heavier than their corresponding slepton counterparts, for the GNLSP case one finds that one
can often get an inversion, i.e., the model gives rise to sleptons heavier than their squark counterparts
within the first and second generations. Specifically, defining 
\beqn
\Delta_{ed}^{(i)}=
2\frac{\left(m_{\tilde d_{1i}}+ m_{\tilde d_{2i}}\right)
-\left(m_{\tilde e_{1i}}+ m_{\tilde e_{2i}}\right)} {  \left(m_{\tilde d_{1i}}+ m_{\tilde d_{2i}}\right)
+\left(m_{\tilde e_{1i}}+ m_{\tilde e_{2i}}\right)},
 ~~i=1,2~,
\label{d-e1}
\eeqn
where $i$ is the generation index, 
for the mSUGRA case one finds that $\Delta_{ed}^{(i)}$  are positive
and typically a significant fraction. However, for the GNLSP case one has 
\be
|\Delta_{ed}^{(i)}| \ll 1,
\label{degen}
\ee
and often $|\Delta_{ed}^{(i)}|$ lie in the range  much smaller than $1\%$. 
Thus the validity of Eq.(\ref{degen}) implies a high degree of degeneracy of the squark and slepton masses for 
the first two generations, and the observation of such a degeneracy will provide
a  strong corroborating evidence along with collider signals for testing 
the validity of the GNLSP models. Of course, a test of  Eq.(\ref{degen}) would require determination of 
the squark and slepton masses with a certain degree of accuracy.  

More generally the lower left hand panel of 
Fig.(\ref{a-tanbeta}) exhibits  $\Delta_{ed}$ as function of
$\Delta_{\rm Co}\equiv \Delta_{\g \na}$  where the gluino NLSP and neutralino LSP coannihilate 
to produce the consistent relic density observations of WMAP. 
The analysis of this  panel exhibits more generally the  results of Table(\ref{sumrule})
in that one
 finds that in all these models 
 $\Delta_{ed}$  is relatively small and often negative. 
 We thus arrive at the important general conclusion that in the model where the gluino is the NLSP 
one gets  a compressed sfermion spectrum\footnote{The compressed sfermion spectrum discussed here is
very different from the one discussed in\cite{SPM}.}
for the first two generations relative to the squarks, with mass differences between
squarks and their slepton counterparts 
which are typically   order a few percent and often less over a wide range  of the parameters space. 
In Fig.(\ref{scaling}) the left panel shows that the  model \a can be discriminated from the bino 
branch of mSUGRA, while   the right panel of Fig.(\ref{scaling})
shows that all GNLSP models can be discriminated from mSUGRA.
Additionally a 
comparison of the right and the left panels also allows a discrimination of \a from  \b.
\begin{table}[htbp]
    \begin{center}
\begin{tabular}{||l||l||c|c||c|}
\hline\hline 
Model & Pattern &    $m_{\tilde e_1} + m_{\tilde e_2}$ & $m_{\tilde d_1} + m_{\tilde d_2}$  &  $\Delta^{(1)}_{de}$ \\
\hline\hline
mSUGRA  & mSP3 & 5377  & 7652  &  35\%\cr
NUSUGRA SU(5)& NUSP13 & 7386 & 7373 & -0.1\% \cr
NUSUGRA SO(10)& NUSP13  & 7369 & 7300 &   -0.1\%\cr
  \hline\hline
\end{tabular}
\caption{Exhibition of the mass compression for sleptons and squarks in the first two generations
in two typical GNLSP SUGRA models with nonuniversalities and a comparison with a mSP3 model point in the mSUGRA case.
  Mass splittings between the sleptons and the squarks are seen to be much smaller for the GNSLP models 
compared to the mSUGRA case.   }
 \label{sumrule}
\end{center}
 \end{table}
 \vspace{-1cm}
\section{Signature analysis at the LHC\label{signature}}
\begin{figure}[h]
  \begin{center}
\includegraphics[width=8.25cm,height=6.75cm]{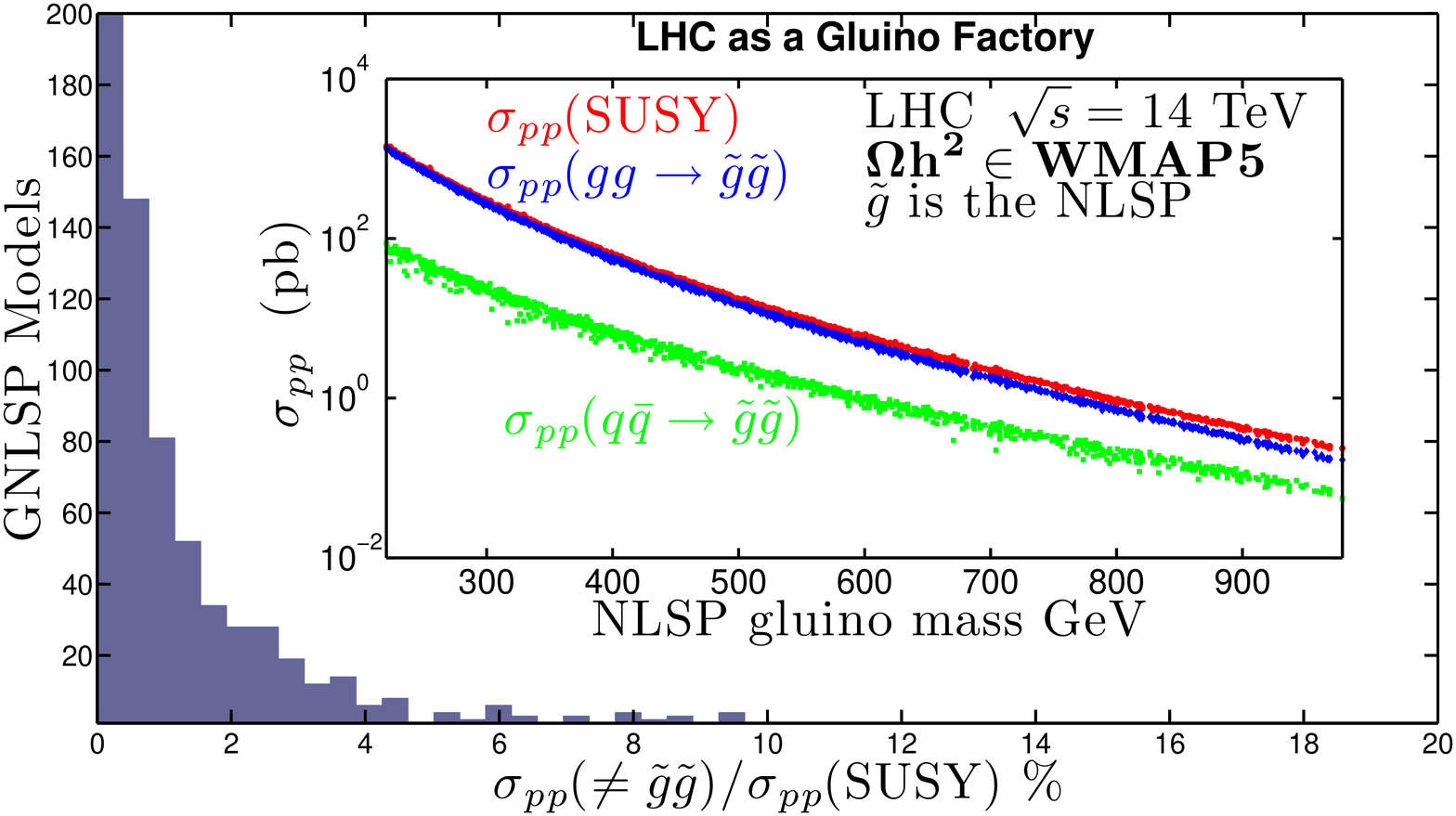} 
\includegraphics[width=8.25cm,height=6.75cm]{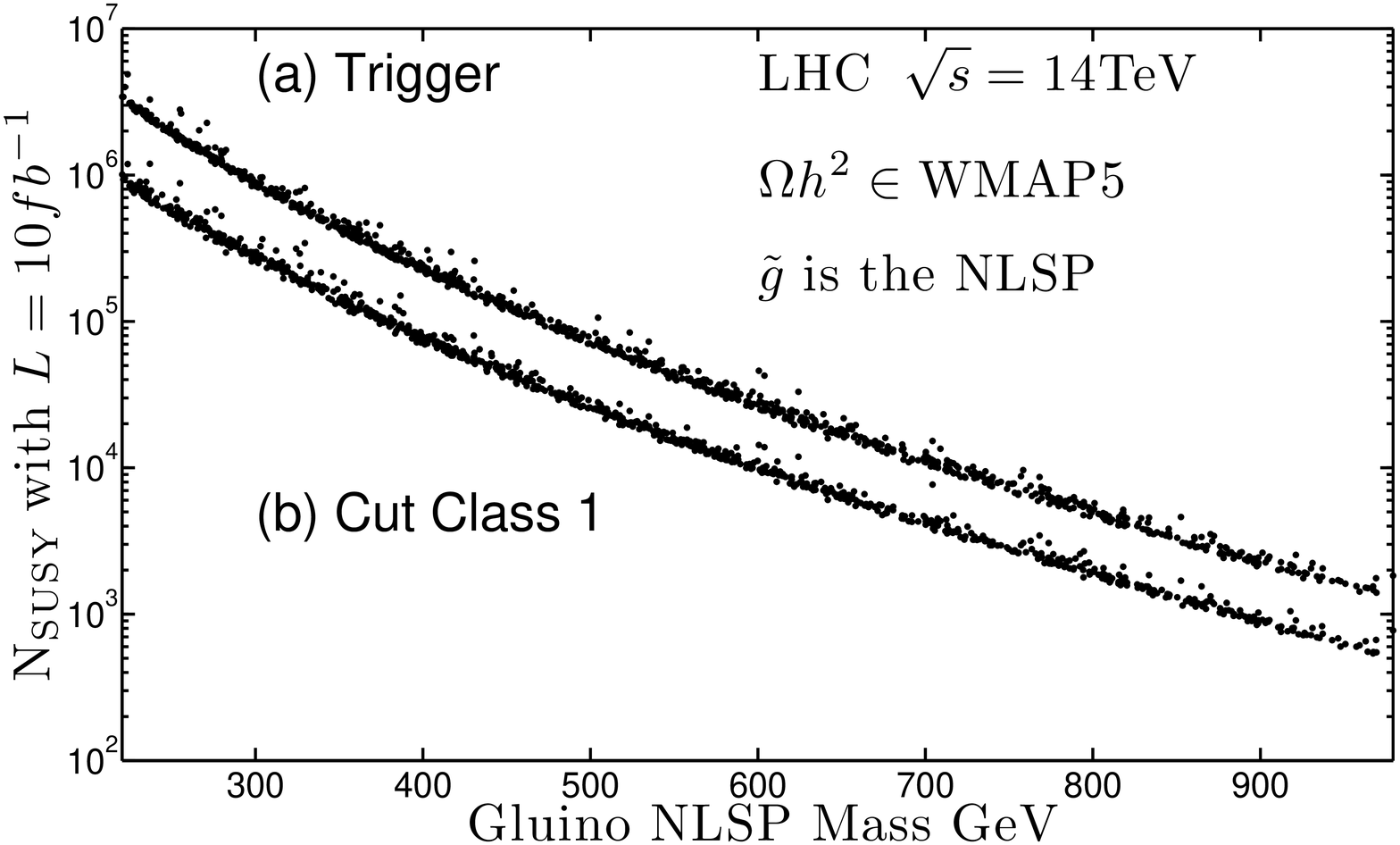} 
\caption{(Color online) Left panel:
Display of $pp$ cross sections including the individual production modes arising from subprocesses $gg\to \tilde g \tilde g$ and  $q\bar q \to \tilde g \tilde g$
and the total SUSY cross sections plotted 
as  a function of the gluino mass showing the dominance of the  $\tilde g \tilde g$ production process.
The analysis of the figures shows  that in the GNLSP case the LHC will turn into a gluino factory.
Right panel: Total number of SUSY events passing the Level 1 (L1) trigger cuts and post trigger level
cuts C1 as defined in the text.   A majority of the
events pass the triggers as can be  seen by comparing with the left and the right panels of Fig.(\ref{gluinofactory1}) and
 taking into account that the right panel of this figure is given for 10 fb$^{-1}$ of luminosity.
}
\label{gluinofactory1}
  \end{center}
\end{figure}

\.4
\noindent
{\bf Gluino Mass and Gluino Production Cross Sections:}
The production cross sections for gluinos were studied early on 
\cite{Kane:1982hw} and the NLO evaluations have also been given\cite{Beenakker:1996ch}.
 A  particularly interesting situation is the one which is discussed in the preceding sections where
 the gluino is the NLSP, as this possibility leads to a rather predictive model. 
 Thus one finds that in the GNLSP case the gluino production cross section dominates all other SUSY processes 
 and further the production is controlled to a large degree 
 by a single process which is $gg\to \tilde g \tilde g$, i.e., 
  $   \sigma_{pp}({\rm SUSY})  \approx \sigma_{pp}(gg \to \g\g) $
where  
 $\sigma({\rm SUSY})$  is the LHC production cross section including all $2\to 2$ SUSY production modes\cite{susy}.
 A numerical analysis of the above is  shown in Fig.(\ref{gluinofactory1}) for the GNLSPs.
  A consequence of the dominance of the processes $gg\to \tilde g \tilde g$ over all others 
  has interesting implications. Specifically it 
   opens up the interesting possibility that a rather precise
  determination of the gluino mass can  be made from a measurement of the production cross section
  of all SUSY processes. 
   Further since the neutralino is linearly related to the gluino through $\Delta_{\tilde g\tilde \chi^0}$ 
  the LHC data in this  case could allow us to determine the neutralino mass with a fair degree of accuracy
should the gluino mass be reconstructed.

It is  worthwhile to pause and comment on the sensitivity of the relic density to 
the codes that are often used for its computation. We exhibit this sensitivity in Table(\ref{GNLSPcompare})
where a small dialing of parameters has been done to keep the relic density in the corridor allowed by WMAP.
The analysis  shows that the sensitivity to the codes is rather small and the
GNLSP model is robust in that  it appears in all the three codes used in Table(\ref{GNLSPcompare}).
 In Table(\ref{GNLSP1}) we compare the leading order (LO)
predictions of PYTHIA and PROSPINO in the GNLSP 
model for the same parameter point given in Table(\ref{GNLSPcompare}),
and show the next to leading order (NLO) prediction using PROSPINO. We also compare the relevant 
branching ratios. 
\begin{table}[h] 
\centering     
\begin{tabular}{cccc}  
\hline              
                      &SuSpect(2.41)\cite{SUSPECT}     & SOFTSUSY(3.0.2)\cite{ben}    & SPheno(2.2.3)\cite{SPHENO} \cr\hline
$\na $               &  336.3                  & 334.5                   &  334.5         \cr
$\g  $                &  382.7                  &  379.8                  &  381.0         \cr
$\cha$                &  424.1                  &  422.4                  &    422.9         \cr
$\ta$                &  451.4                  &  464.7                  &     447.4      \cr
$(\Omega h^2)_{\na}$          &  0.115                  &  0.105                  &   0.117       \cr
$(\delta_2,\delta_3)$ & (-0.340, -0.835)         & (-0.340, -0.824)           &  (-0.340, -0.823)        \cr \hline
\end{tabular} 
\caption{A comparison of the sparticle spectra and of the relic density for a GNLSP model
with  $(m_0, \mh, A_0, \tan \beta) =(1387, 792, 3026, 27) $ (all masses in $\GeV$), with $\mu > 0$ 
for $m_{t} = 170.9 \GeV$
and $\delta_2 =-0.340$ 
with $m_b(m_b)$ and $\alpha_s(M_Z)$ taken with the default values of MicrOMEGAs (MO).
For these models $\cha \lesssim \nb$ (NUSP13). 
We use MO 2.2CPC for the first two cases and MO 2.07 for SPheno. This particular comparison is made at the  
perturbative level.} 
\label{GNLSPcompare} 
\end{table}
\begin{table}[h] 
\centering     
\begin{tabular}{|c|}  
\hline              
PYTHIA\cite{PYTHIA}  $ {\rm \sigma_{\rm LO}}$ \\
\hline
$g  g  \to  \tilde g  \tilde g  =70\pb$                       
\\$q  \bar q \to  \tilde g  \tilde g  =6.2\pb$   \cr               
$q_j  g  \to  \tilde{q}_{jL}  {\tilde g} =1.1\pb $             
\\ $g  g  \to  {\tilde t}_1  {\bar {\tilde t}}_1 =1.5\pb  $  \cr    
$\rm{else} \ll 1\pb$  \cr 
\hline
$\sigma^{\rm LO}_{\rm SUSY} = 80\pb $  \cr                       
\hline  
\end{tabular} 
\begin{tabular}{|c|c|c|}  
\hline              
PROSPINO\cite{PROSPINO}   $ {\rm \sigma_{\rm LO}}$ & $K_{\rm NLO}$  & ${\rm \sigma_{\rm NLO}}$ \\
\hline
$\g  \g  =84.3\pb$                   &  1.72    &   $145 \pb $\cr
$\tilde{q}  {\tilde g} =3.12\pb $ &   1.60    &   $5.0 \pb $   \cr 
${\tilde t}_1  {\bar {\tilde t}}_1=0.80\pb $ &   1.55    &   $1.24 \pb $   \cr 
$\rm{else} \ll  1\pb$                &     -    & -\cr                
\hline
 $\sigma^{\rm LO}_{\rm SUSY} = 88.5\pb $ & $\Longrightarrow $ & $\sigma^{\rm NLO}_{\rm SUSY} = 151.6\pb $
\cr                       
\hline  
\end{tabular} 
\begin{tabular}{|c |c|c|}  
\hline             
Decay &  $ {\rm BR } $ PYTHIA\cite{PYTHIA} &  $ {\rm BR } $ SUSY-HIT\cite{susyhit}\\
\hline
$ \g \to (b \bar b \na   ,  u \bar u \na ,  d \bar d \na) $      &  (20,61,19)\%  &  (20,61,19)\%  \cr  
     $   \g \to \na g    $                             &    -              &   0.03\%   \cr
$ \tilde q_L \to \g (u,d)_L $      &  82\%   &  86\%  \cr 
$   \ta \to \chan^{+} b    $ &  100\%           &  100\%  \cr  
\hline  
\end{tabular} 
\caption{
A specific exhibition of the dominance of the process $g g\to \tilde g\tilde g$ in $pp$ collisions at LO and NLO for the GNLSP 
model point given in Table(\ref{GNLSPcompare}). 
}
\label{GNLSP1} 
\end{table} 
\begin{figure}[h]
  \begin{center}
\includegraphics[width=8.25cm,height=6.5cm]{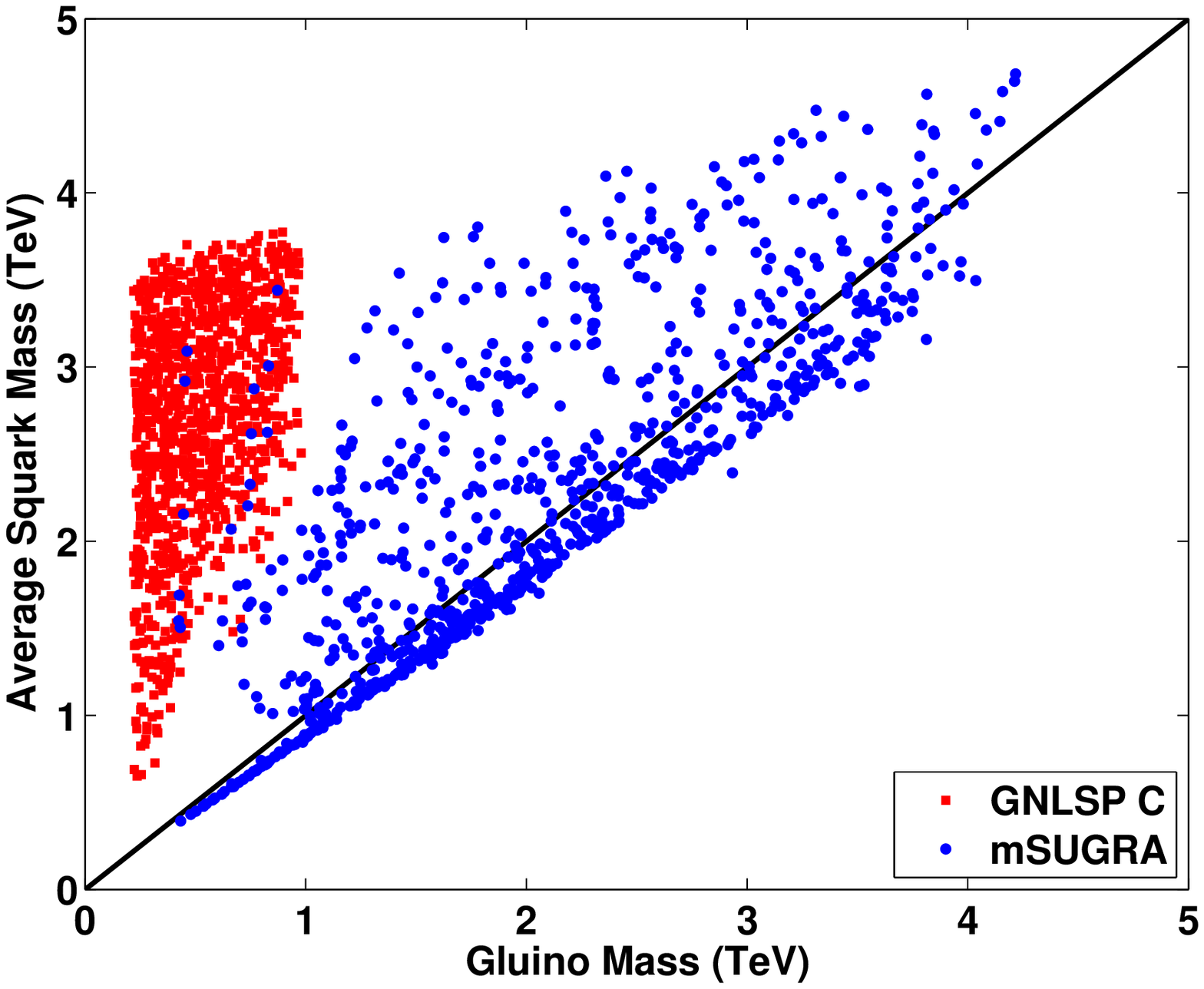} 
\includegraphics[width=8cm,height=6.5cm]{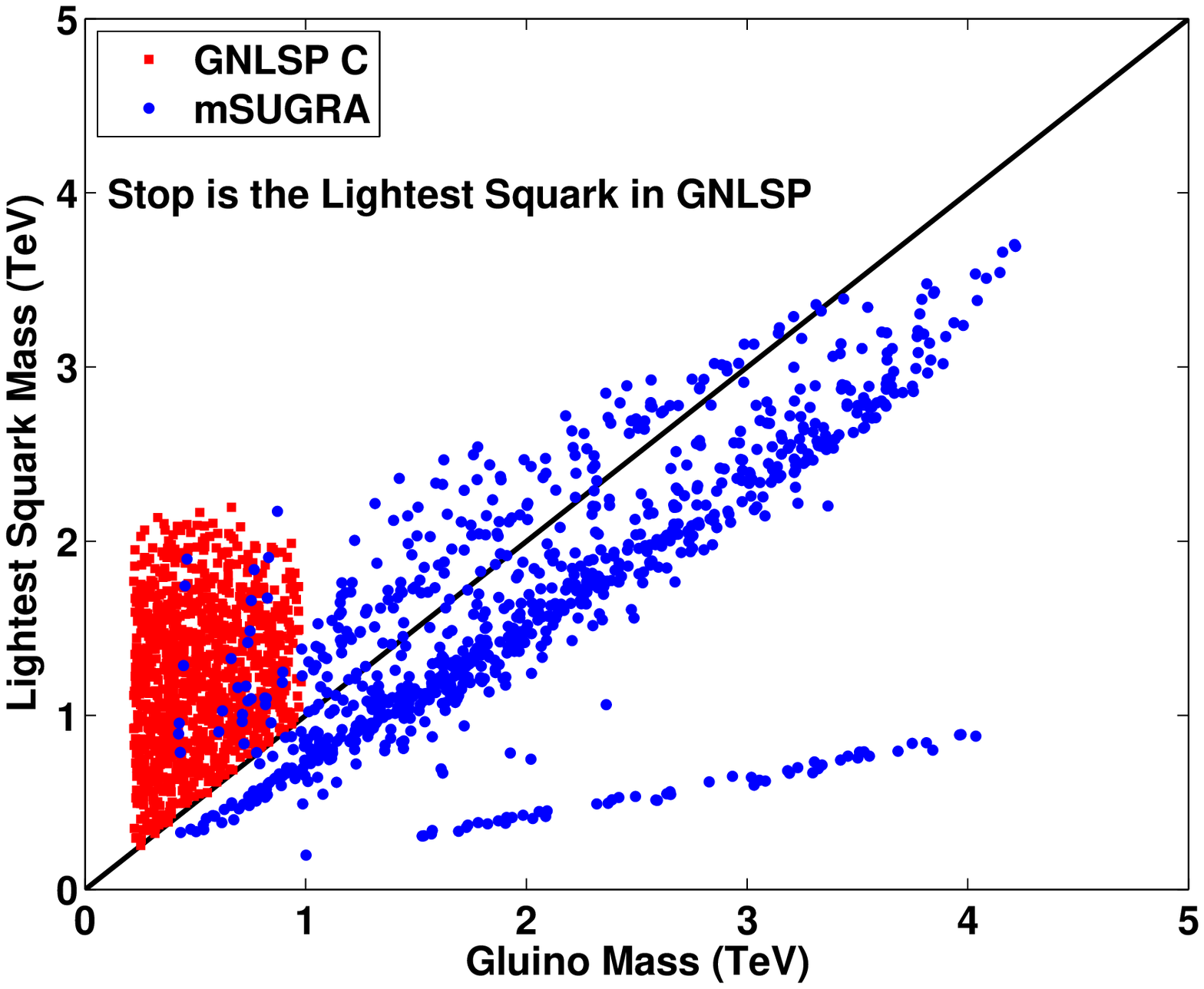} 
\caption{(Color online)
Left panel: A display of the average squark mass vs the gluino mass in GNLSP$_{\rm C}$ and a comparison with mSUGRA. Right panel:
A display of the lightest squark mass
  vs the gluino mass in GNLSP$_{\rm C}$ and a comparison with mSUGRA.
The gluino mass is relatively light in the GNLSP models with an upper limit of a TeV under the 
assumed naturalness  assumptions. The lightness of the gluino mass leads to the dominance  of the 
$\tilde g \tilde g$ over all other sparticle production processes in the GNLSP models.
}
\label{mass}
  \end{center}
\end{figure}
One
observes that the squarks decay back into a gluino  with a large branching ratio. 
The  NLO calculation retains the dominance of the $\sigma^{\rm NLO}_{\g\g}$ at the level of 96\%, 
i.e., $\sigma_{\tilde g\tilde g}^{\rm NLO}= 0.96 ~\sigma_{\rm SUSY}^{\rm NLO}$. 
We note that the model in Table(\ref{GNLSP1}) has a very small branching fraction  $   \g \to \na g    $ 
(calculated with SUSY-HIT\cite{susyhit} and a very similar suppression is seen with ISAJET \cite{Paige:2003mg} ($<$0.1\%)).
This is to be contrasted with the  
 $\rm GNLSP_{\rm Co}$ model (see Eq.(\ref{co})), which has a $\g \to \na g$ of 65 \%  with SUSY-HIT and 89\% with ISAJET. 
The gluino decay branching ratios are discussed in further detail later. 

In the left panel of Fig.(\ref{mass}) we give a comparison of the average mass of the squarks
 vs the gluino mass in GNLSP  and compare it to that for the mSUGRA model. 
 A similar analysis for the lightest squark mass is given on the right panel. 
In both cases one finds 
that aside from a very small region, the spectra from these two models do not overlap. Quite remarkably for 
the assumed naturalness conditions as described in Sec.(\ref{parameter}) the gluino mass in the GNLSP models  has an upper limit 
of about a TeV, while in mSUGRA this limit extends far  beyond. On the other hand, the upper limit on the squark masses
is quite large extending to several TeV in both cases. It is the relative lightness of the gluino mass in the GNLSP case
that enhances the production of the gluinos relative to all other  SUSY production processes such as $\tilde g\tilde q$ and
$ \tilde t_1  \bar{\tilde  t}_1$ at the LHC. This further explains the result of 
Fig.(\ref{gluinofactory1}) and of Table(\ref{GNLSP1}) which show that the $\tilde g \tilde g$ production cross section
dominates  over all others. Thus in effect, in the GNLSP case, the LHC will become a gluino factory.  Nonetheless,
the production cross section for the squarks is still significant and their production could be detectable with
efficient cuts. 
For example, the stop
is usually the lightest squark in GNLSP models, so its production could be significant, and it
can decay via $\tilde t_1\to  \tilde \chi^{+} + b \to W^+ + b + \tilde\chi^0$.
Since the $W^+$ has decays into $l^++\nu$,
 the stop will have some leptonic signatures while the gluino is a pure jet signal. An analysis of these  leptonic
signals requires a further dedicated analysis which is left for a future analysis.  
We also note that superficially the overlap of the few GNLSP points with the mSUGRA point might  be  construed 
as the existence of a degeneracy for these parameter points. However, we need to keep in mind that what is  plotted
is an average squark mass. Further, as  already discussed there is a significant splitting between the squark masses and the 
slepton masses for the mSUGRA model for the first  two  generations while there is very little splitting in this case
for the GNLSP model. Thus at the very least the slepton-squark splittings lift any perceived degeneracy indicated
 in Fig.(\ref{mass}).
We should also point out that although the production cross section of squarks and gluinos can
be comparable for GNLSP models and for mSUGRA models, their LHC signatures tend to be
significantly different due to the mass spectra of sparticles being rather different and specifically this is so 
since the gluino is the NLSP in the class of models we consider.

\noindent
{\bf Early LHC Discovery Prospects at Low Luminosity:}
Each GNLSP model is subject to the experimental constraints
as discussed in Section(\ref{exp}). We investigate the LHC signatures of
1070 such GNLSP
models
with PYTHIA coupled to  PGS4 \cite{PGS}.
Branching fractions are computed with SUSY-HIT and fed directly into the
PYTHIA decay table via the SUSY
Les Houches Accord (SLHA) \cite{SKANDS} interface.  For the GNLSP models, this is quite important
as one must take into account the radiative decay $\g \to \na g$ which
can be substantial
for this class of models. The LHC detector simulation proceeds
with PGS4  with the Level 1 (L1) triggers designed to mimic
the Compact Muon Solenoid detector
(CMS) specifications \cite{Ball:2007zza} with the LHC detector card.
Specifically the L1 trigger level cuts that
are imposed are as follows\cite{PGS}:
(1)  inclusive isolated lepton ($\mu$/$e$)    (30 GeV);        
(2)  lepton plus jet   (20 GeV, 100 GeV);  
(3)  isolated di-leptons  (15 GeV);       
(4)  di-leptons plus jet   (10 GeV, 100 GeV);  
(5)  isolated di-leptons  (10 GeV);  
(6)  isolated lepton plus isolated $\tau$ (15 GeV, 45 GeV);  
(7)  isolated di-tau   (60 GeV);           
(8)  inclusive isolated photon   (80 GeV);         
(9)  isolated di-photon    (25 GeV);         
(10) inclusive $\pt$   (90 GeV);           
(11) inclusive single-jet    (400 GeV);          
(12) jet plus $\pt$  (180 GeV, 80 GeV);  
(13) acoplanar jet and $\pt$ with $(1 < \Delta \phi < 2)$   (100 GeV, 80 GeV)   (jet, $\pt$);  
(14) acoplanar di-jets with ($\Delta \phi  < 2$)  (200 GeV).       
 Muon isolation is controlled by employing the cleaning script in PGS4.
SM backgrounds have been generated with
QCD multi-jet production due to light quark flavors,
 heavy flavor jets ($b \bar b$,  $t \bar t$), Drell-Yan,
single $Z/W$ production in association with quarks and gluons ($Z$/$W$+ jets),
and  $ZZ$, $WZ$, $WW$ pair production.
The standard criteria for the discovery limit are imposed, namely, that the SUSY signal
is taken as discoverable if the number of SUSY events exceeds 
 $5\sqrt{N_{\rm SM}}$ or 10 whichever is larger,  i.e.,
${\rm N_{SUSY}}>{\rm Max}\left\{5\sqrt{N_{\rm SM}},10\right\}$.
We implement several classes of post trigger level cuts to analyze the SM background and 
our event samples. 

\vspace{-.1cm}
\noindent
{\bf Post Trigger Level Cuts:}
As already noted the dominant sparticle production process at the LHC will be $gg\to \tilde g \tilde g$,
and the dominant decays of the gluino are  $q\bar q \na$ and depending
on the particular part of the parameter space the decay $\tilde g\to g \tilde\chi^0$\cite{Haber:1983fc,wells,Baer:2006dz} can also be large,
and in fact can dominate over the 3 body decay, depending on a confluence of the 
following:
(a) the lightness of the
gluino, (b) the largeness of the squark masses, and  (c) the neutralino mixing matrix.
While a heavy gluino can decay into quarks and squarks and
into weak gauge bosons $W^{\pm}, Z$\cite{Baer:1987ye}, these decays are either kinematically forbidden,
such as into quarks and squarks, or are highly suppressed for the GNLSP models we are discussing here.
 The above implies that  GNLSP models will lead to  a preponderance of jet signatures.
In our analysis we impose three different sets of post trigger level cuts to 
optimize the signal and reduce the background from the standard model processes.
We classify these as: (i) Class 1 (C1), (ii) Class 2 (C2), and 
(iii) Class 3 (C3) post trigger level cuts. We discuss these in some detail below and discuss the signals
that are best detected with these three classes of cuts. 
Before proceeding further, we note that 
the missing 
transverse momentum is an important cut both as a  trigger level as well as 
a post trigger level cut allowing one to increase the
signal relative to the background.  
Thus SUSY models with  a LSP which is massive tend to produce events at
the hadron colliders with a larger missing energy. In order to suppress
the Standard Model background, 
usually a  large missing transverse momentum cut
 is employed. We note further that since $b$ and $\bar b$ are produced
in the processes $gg\to q\bar q,q\tilde g, \tilde g \tilde g$ as well as with a certain
fraction in $gg\to \tilde g \tilde g$ with 
subsequent decays of $\tilde g$  (this latter case being the most relevant one discussed here),
 one has a significant number of b jets 
 produced in these events. Thus b -tagging is 
 a useful instrument in their identification.
These 
features will be seen in the post trigger level cuts we discuss below.


\noindent
{\em 
(i) Class 1  Post Trigger Level Cuts (C1)}
\.4
\begin{enumerate}
\item Electrons, and muons with $P_T>10$ GeV (where $P_T$ is the transverse momentum)
and $|\eta|<2.4$ (where $\eta$ is pseudorapidity) are selected. 
\item Jets with $P_T>60$ GeV and $|\eta|<3$ are selected.
\item Events with $\pt>200$ GeV are selected. 
\end{enumerate}
\.4
Since the GNLSP models have a gluino which lies close to  the LSP consistent 
with the relic density constraints, 
a large portion  of events generated by the SUSY processes have a less energetic jet signal and a
relatively smaller missing energy than may be observed, for example, in the case
of stau 
coannihilation (see \cite{Feldman:2008en} for analysis of such a signature). 
Taking the above into account we modified the imposed C1 post trigger level cuts in 
C2 and C3 which are to be discussed below.
As already emphasized, the SUSY production cross section are dominated by gluino 
pair production and decays of the gluino. Therefore, we only select events 
that have at least two jets that pass our jet selection condition.  We also investigate
the effects of putting a softer missing $P_T$ cut and jet $P_T$ cut. 
\begin{figure}[t]
\vspace{1.5cm}
  \begin{center}
\includegraphics[width=7.5cm,height=7cm]{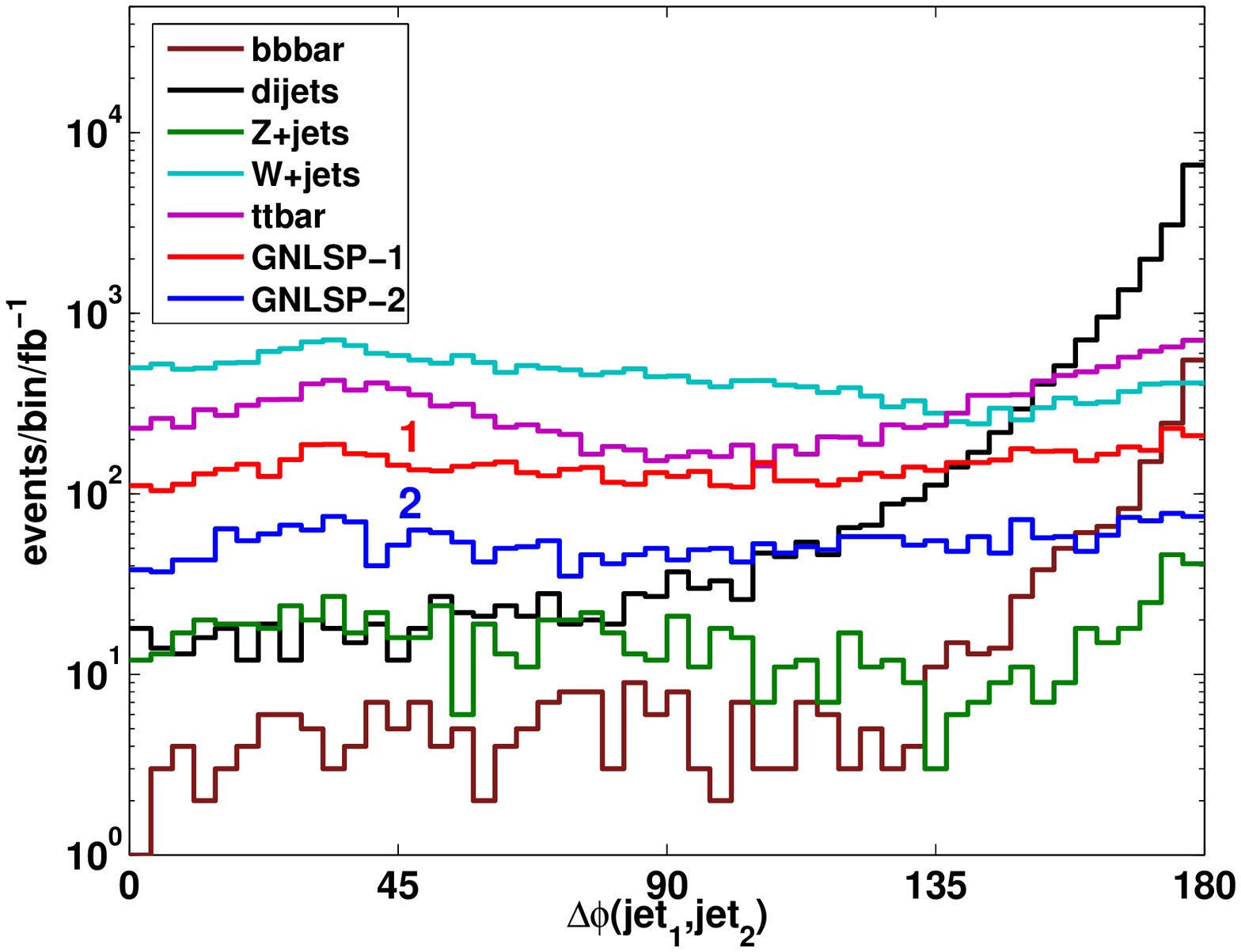}  
\includegraphics[width=7.5cm,height=7cm]{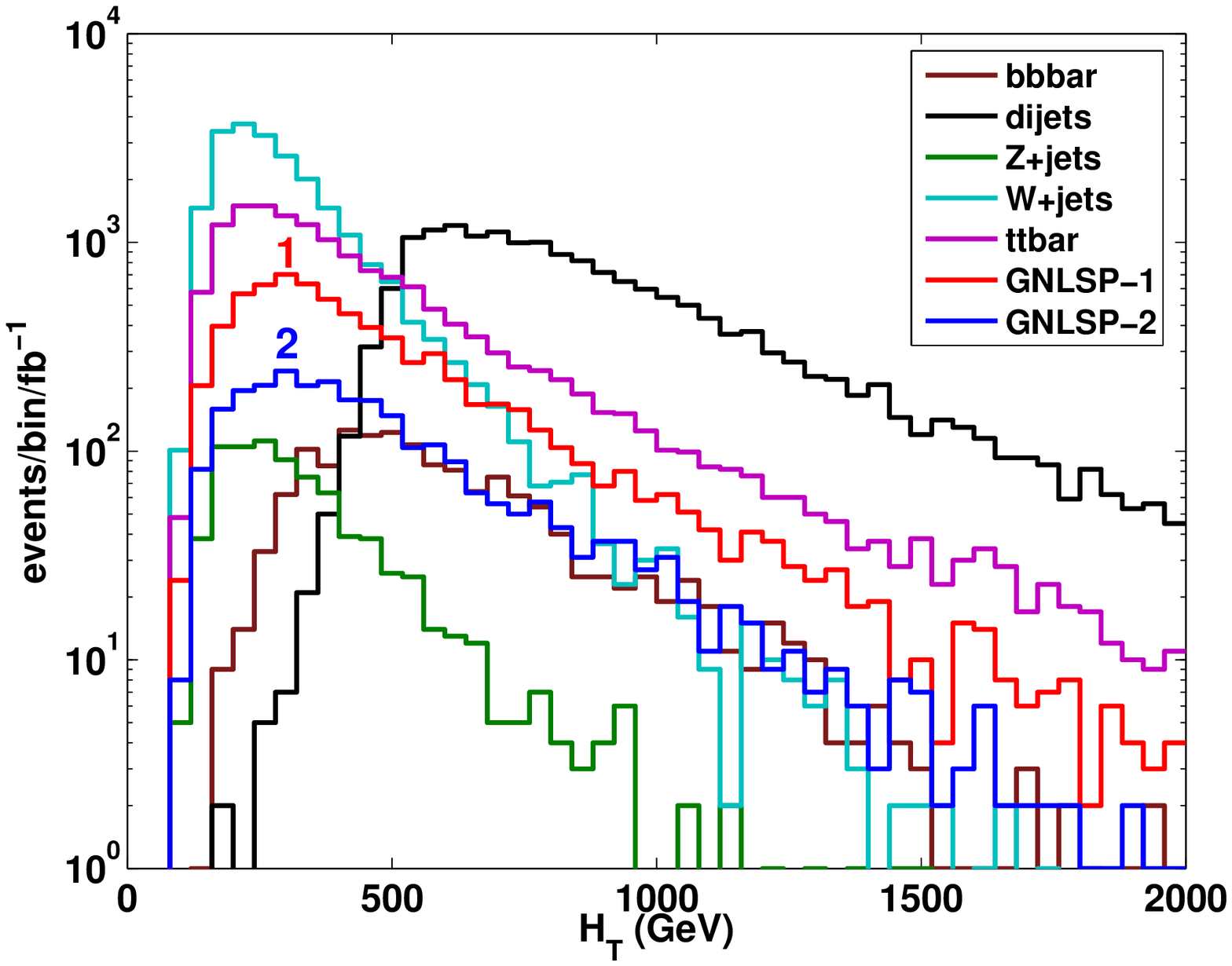}
\caption{(Color online) Left panel: An analysis of events/bin/fb$^{-1}$ as a function of the azimuthal angle $\Delta \phi({\rm jet}_1~{\rm jet}_2)$ between
the two hardest jets in the GNLSP model relative to the SM backgrounds with C2 cuts. Right panel: 
$H_T=\sum_{jets} P_T$ 
distributions of GNLSP models and SM backgrounds with C2 cuts. These distributions act as a guide for implementing the C3 cuts as discussed in the text. }
 \label{distributions}
  \end{center}
\end{figure}

\noindent
{\em (ii) Class 2  Post Trigger Level Cuts (C2)}
\.4
\begin{enumerate}
\item Electrons, and muons with $P_T>10$ GeV and $|\eta|<2.4$ are selected. 
\item  Jets with $P_T>50$ GeV and $|\eta|<3$ are selected.
\item Events with $\pt >150$ GeV are selected. 
\item Events with at least 2 jets are selected. 
\end{enumerate}
\.4

The dominant SM background for GNLSP models are from QCD, $Z$/$W$+ jets, $b \bar b$, and $t \bar t$
and we focus on these in Fig.(\ref{distributions}).
In the left panel of Fig.(\ref{distributions}) we give an analysis of these backgrounds at the LHC with 
 events/bin/fb$^{-1}$ as a function of the  azimuthal angle $\Delta\phi ({\rm jet}_1, {\rm jet}_2)$ between the two leading jets.
 We also show the distributions for 
 two GNLSP  model points with masses of 400 and 500 GeV respectively. 
A similar analysis is presented but as a function of the total jet $P_T$ (labeled $H_T$) in the
    right panel of Fig.(\ref{distributions}), where again we also exhibit the distributions for two 
  GNLSP model points. 
   The analysis provides a good starting point for charting out the post trigger level
 cuts needed to  reduce the background and enhance the signal.   
Thus the analysis suggests that one may cut out the events that have a large $\Delta\phi ({\rm jet}_1, {\rm jet}_2)$
as this cut will suppress the 
QCD background due to light quark flavors, $b \bar b$ and $t\bar t$. 
A veto on isolated electrons or muons was applied to 
reject the background events containing $W$ or $Z$ leptonic decays. A simple counting of events 
after applying C2 cuts reveals that the GNLSP models are nearly lepton free, as is the case for the $b\bar{b}$ 
and di-jets background. However, the standard model backgrounds due to Drell-Yan process
 and ($Z$/$W$+ jets) have ($31\%\sim 36\%$) 
of the total events that contain electrons or muons, while for $t \bar t$ background it is about $45\%$. 
The  $ZZ$, $WZ$, $WW$ pair production resulting in multi-leptonic backgrounds 
have an even larger percentage of leptonic events. Therefore,  the $e/\mu$ veto significantly enhances the 
GNLSP signals over the standard model background.  This leads us to investigate  cut class C3. \\\\

\noindent
{\em (iii) Class 3 Post Trigger Level Cuts (C3)}
 \.4
\begin{enumerate}
\item Apply cut set C2. 
\item Electron or muon veto is imposed.
\item $H_T\equiv\sum_{jets}P_T>400$ GeV.
\item  The azimuthal angle $\Delta \phi({\rm jet}_1,{\rm jet}_2)$ between jet1 (the hardest jet) and  jet2
(the second hardest jet)
is chosen so that $\Delta \phi({\rm jet}_1,{\rm jet}_2)<3\pi/4$.
\item The azimuthal angle 
$\Delta\phi({\rm jet}_1,\pt)$ between jet1 (the hardest jet)  and $\pt$ 
is chosen so that $\Delta \phi({\rm jet}_1, \pt)> \pi/2$. 
\item The azimuthal angle 
$\Delta \phi({\rm jet}_2,\pt)$ between jet2 (the second hardest jet) and $\pt$ 
is chosen so that $\Delta \phi({\rm jet}_2,\pt)> \pi/4$. 
\end{enumerate}
As is indicated from the preceding discussion, jets with and without tagged b jets 
are important signals for the discovery of the GNLSP models. Another important signature
is  $\langle \pt \rangle $, which is the average magnitude of the missing transverse momentum,
where the average extends over all events passing the cuts.
\begin{figure*}[t]
  \begin{center}
  \includegraphics[width=7.5cm,height=7cm]{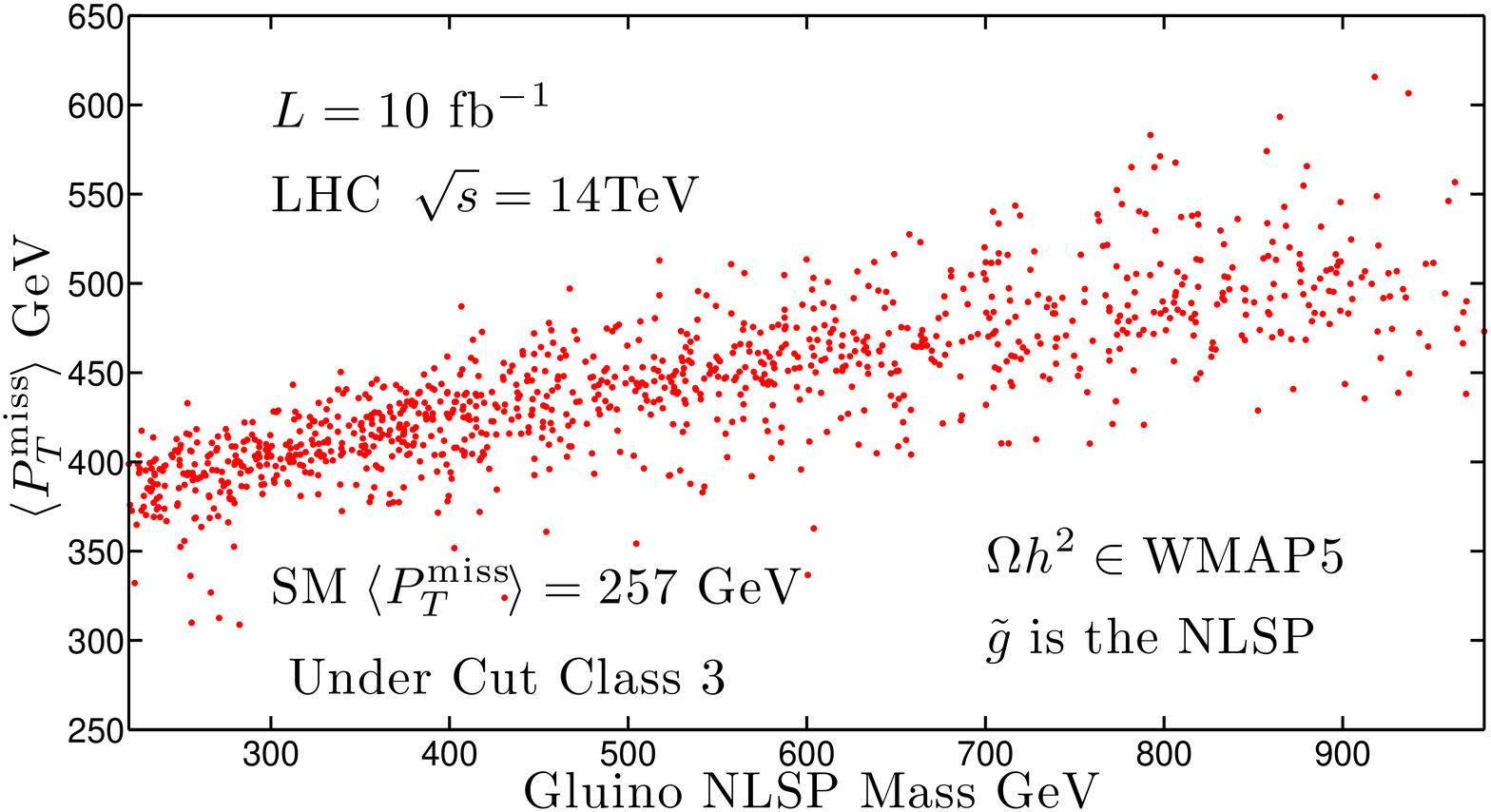}
\includegraphics[width=7.5cm,height=7cm]{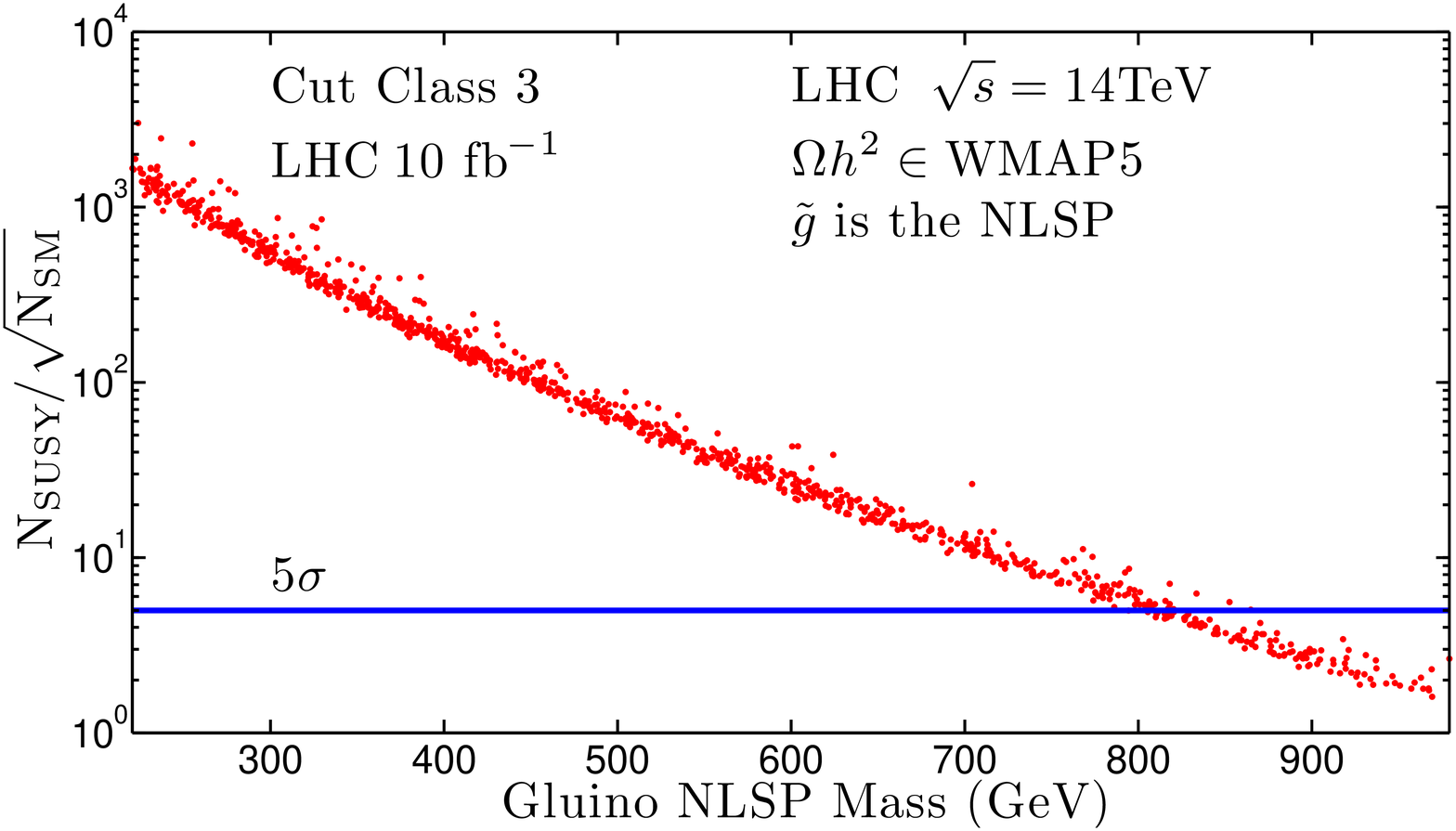}  
\caption{(Color online) The analysis above  is with post trigger level cuts C3 and with an integrated luminosity of 10 fb$^{-1}$.
Left panel: average $\pt$ vs the gluino mass; Right panel: the discovery reach in
SUSY events  vs the gluino mass. 
}
\label{C3cuts}
  \end{center}
\end{figure*}
 We discuss now several signatures for the GNLSP models as
given in Fig.(\ref{C3cuts}). The analysis of Fig.(\ref{C3cuts}) is given at $\sqrt{s} = 14~{\rm TeV}$ under 
 post trigger level cuts C3 with an  integrated luminosity of 10 fb$^{-1}$.
The  left panel of Fig.(\ref{C3cuts}) exhibits  the  average $\pt$ as a function of the gluino mass.
 Here one finds that essentially all of the parameter points of the GNLSP models have an  
 average $\pt$ which is larger, and often significantly  larger, than for the standard model case for which
 the average $\pt$ is found to be $\sim$ 257 GeV under the same set of cuts. The  right panel illustrates the 
ratio $N_{\rm SUSY}/\sqrt{ N_{\rm SM}}$ vs the gluino mass where $N_{\rm SUSY}$ is the 
 total number of  SUSY events and $N_{\rm SM}$ is the standard model background,
again with the imposition of post trigger level cuts C3. Here one finds that the ratio  
$N_{\rm SUSY}/\sqrt{ N_{\rm SM}}$ lies above 5$\sigma$ discovery limits for gluino masses up to 800 GeV making this 
ratio an important channel for the discovery of the GNLSP models.

\begin{figure*}[t]
  \begin{center}
\includegraphics[width=7.5cm,height=7cm]{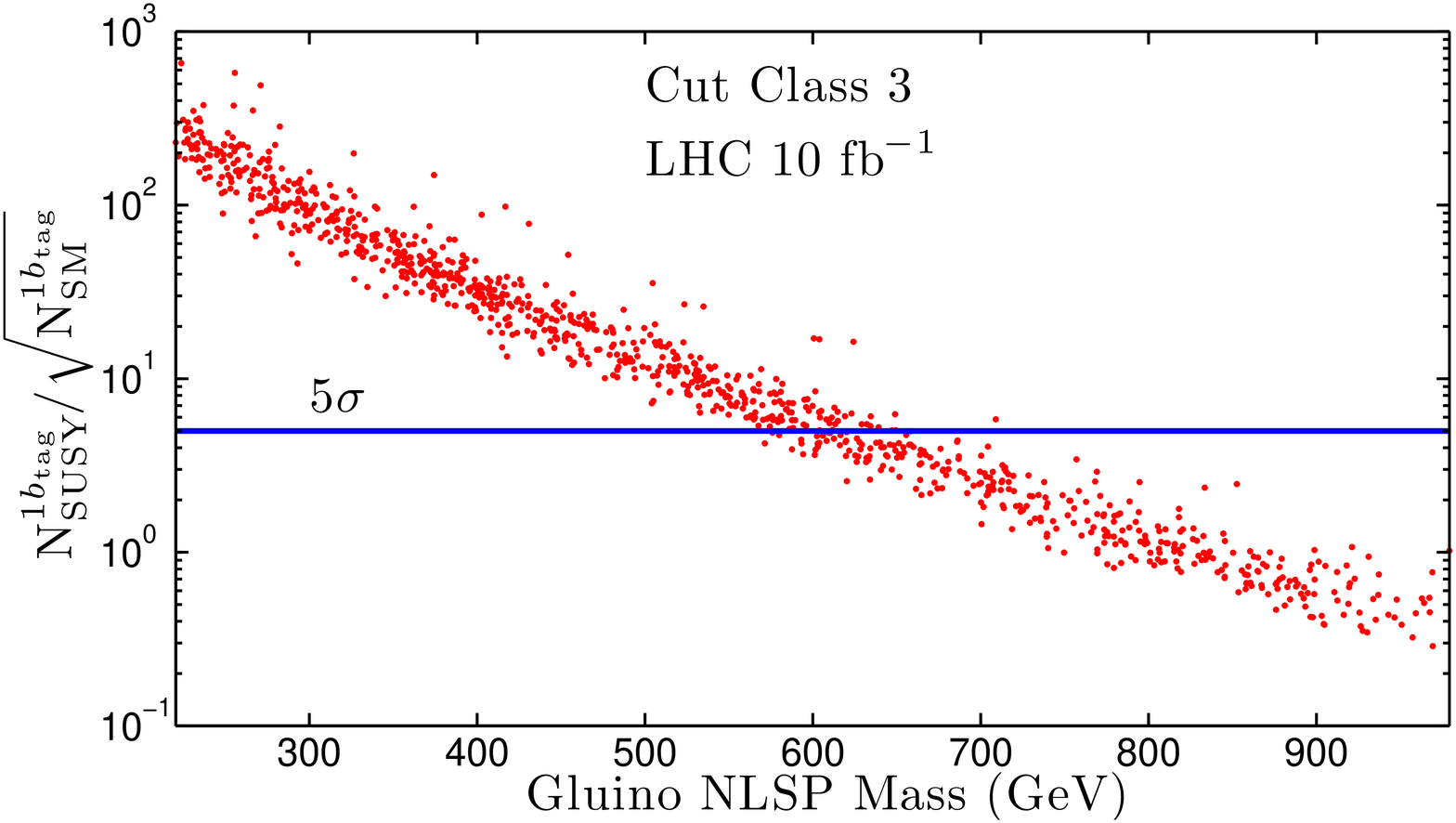}
  \includegraphics[width=8cm,height=7cm]{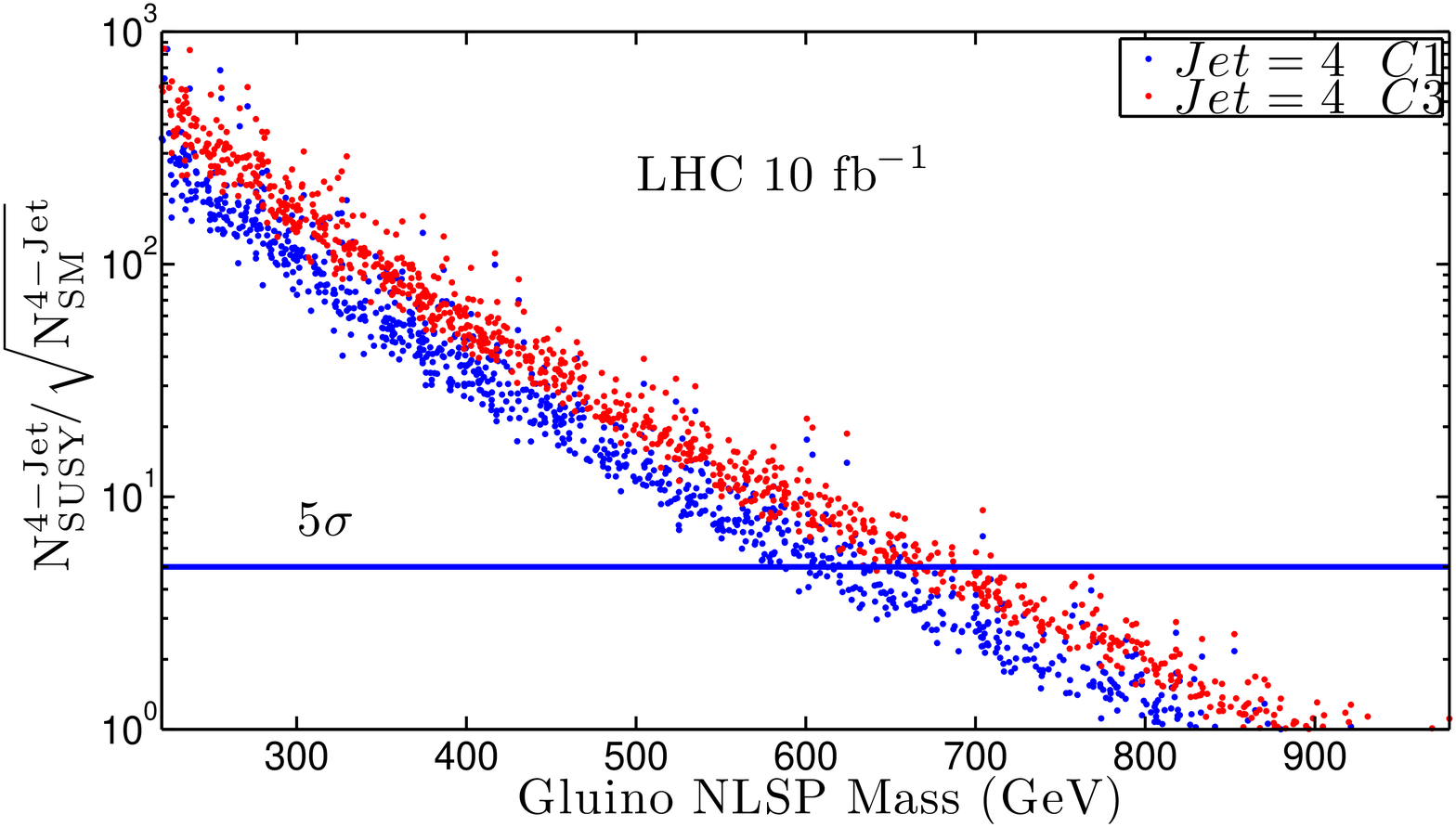}
\caption{(Color online) All Cuts and constraints as in Fig.(\ref{C3cuts}):
Left panel: the discovery reach with 10/fb for SUSY events with 1 tagged b-jet vs the gluino mass, and
Right panel: A comparison of post trigger level cuts C1 and C3 for the discovery of a 4-jet signal with 
an integrated luminosity of 10fb$^{-1}$ at the LHC. 
It is  seen that the specialized cuts C3 enhance the 4J reach by roughly 100 GeV relative to the
post trigger level cuts C1. The C3 cuts reduce the 
 SM background while  allowing a large n-jet signal  for $n\geq 2$ with the
largest signal to background enhancement appearing  for $n=4$.
}
\label{C3cutschannelandcompare}
  \end{center}
\end{figure*}
 The left panel of Fig.(\ref{C3cutschannelandcompare})  demonstrates that 
the  events containing  a single tagged b-jet leads to a discoverable signal over a wide range of gluino masses. 
Specifically this panel gives the 
ratio $N_{\rm SUSY}^{1 b_{jet}}/\sqrt{ N_{\rm SM}^{1b_{jet}}}$ vs the gluino mass 
where $N_{\rm SUSY}^{1 b_{jet}}$ is the 
  number of  SUSY events with 1 tagged b-jet and
 $N_{\rm SM}^{1b_{jet}}$ is the number of standard model events with 1 tagged b-jet  under the
 imposition of class C3 cuts. This channel provides a $5\sigma$ discovery for gluino masses up to about
 600 GeV. 

We briefly comment on the ability to tag b-jets in this case. 
Naively one would expect that for cases in which the dominant decays
 are $\g \to \na q \bar q$, with sizeable branchings
in b quarks, that the b-jet events would be a gold plated signal. However, here the b's
come out rather soft since the mass splitting of the $(\g -\na)$ is rather small, 
typically around 50 GeV (the phenomenon responsible for the satisfaction
of the relic density is precisely this mass split). While the ability to tag the b's
is possible, it is indeed more difficult than the canonical situation seen on the Hyperbolic
Branch of REWSB \cite{hb} (for recent work on b-tagging analyses see \cite{Baer:2007ya,Kadala:2008uy,Feldman:2008jy}).

Nonetheless,  as discussed above the signal for singly tagged b-jets
is strong enough that it can be a useful one. 
 The right panel of  Fig.(\ref{C3cutschannelandcompare}) we compare the 4-jet discovery limits under C1 and C3 cuts. Here it
is found that the SM background drops by an appreciable amount as one goes from C1 cuts to C3 cuts, allowing one to extend the discovery
limit in 4-jet channel by over 100 GeV with C3 cuts relative to imposing the C1 cuts in this channel.
Amongst the classes of n-jet events, we find this channel to be the most enhanced 
when passing from C1 to C3. 
Combining all the channels analyzed in Fig.(\ref{C3cuts},\ref{C3cutschannelandcompare}) one finds 
 that 
gluino masses up to 800 GeV are discoverable by this technique for a  GNLSP model
with just 10 fb$^{-1}$ of integrated luminosity.
Thus the validity of the GNLSP models can be tested with first data from the LHC. 

\section{Dark matter detection in the GNLSP models\label{dark}}
\begin{figure*}[h]
  \begin{center}
\includegraphics[width=8cm,height=6.5cm]{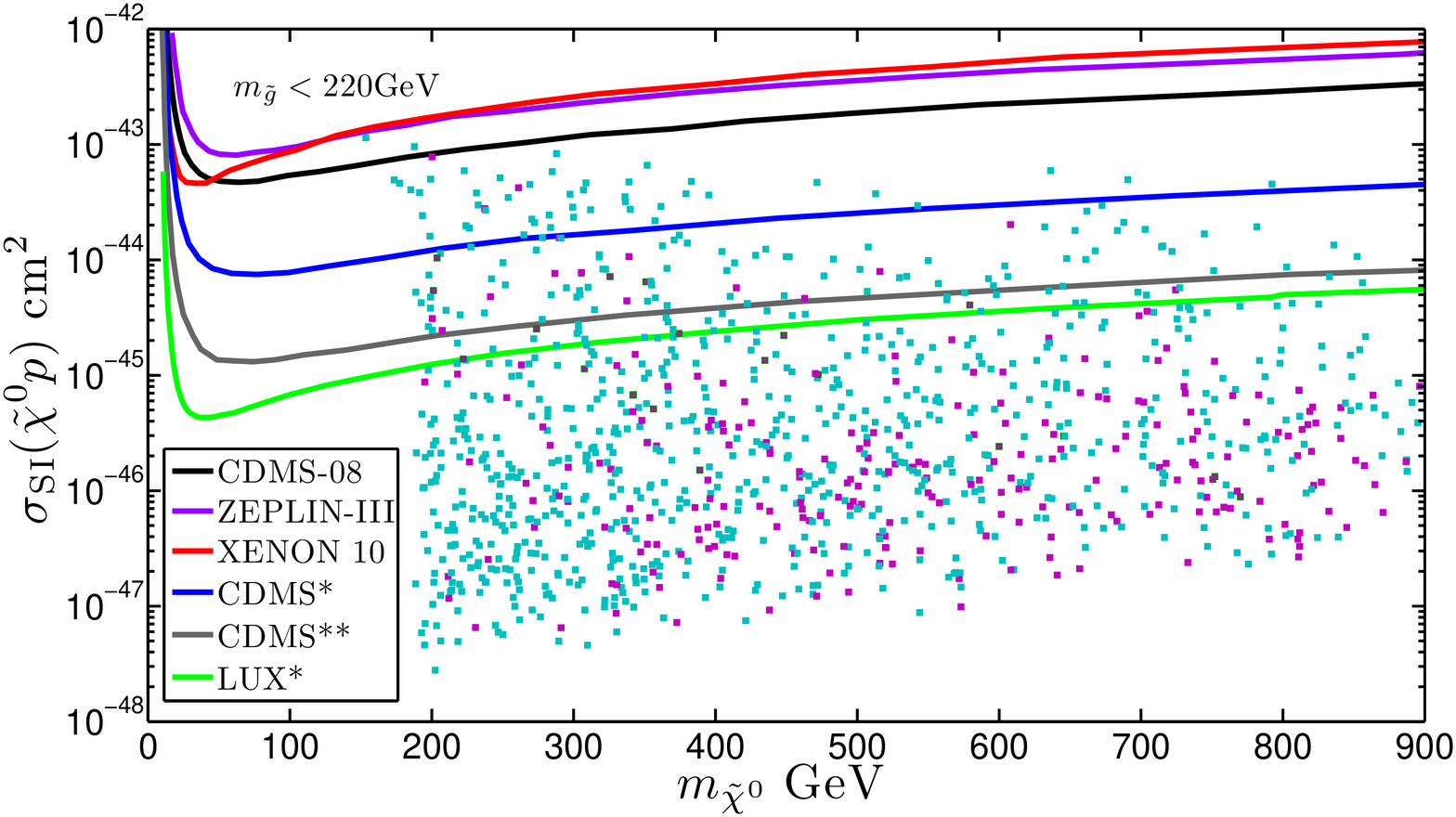}
\includegraphics[width=9.0cm,height=6.5cm]{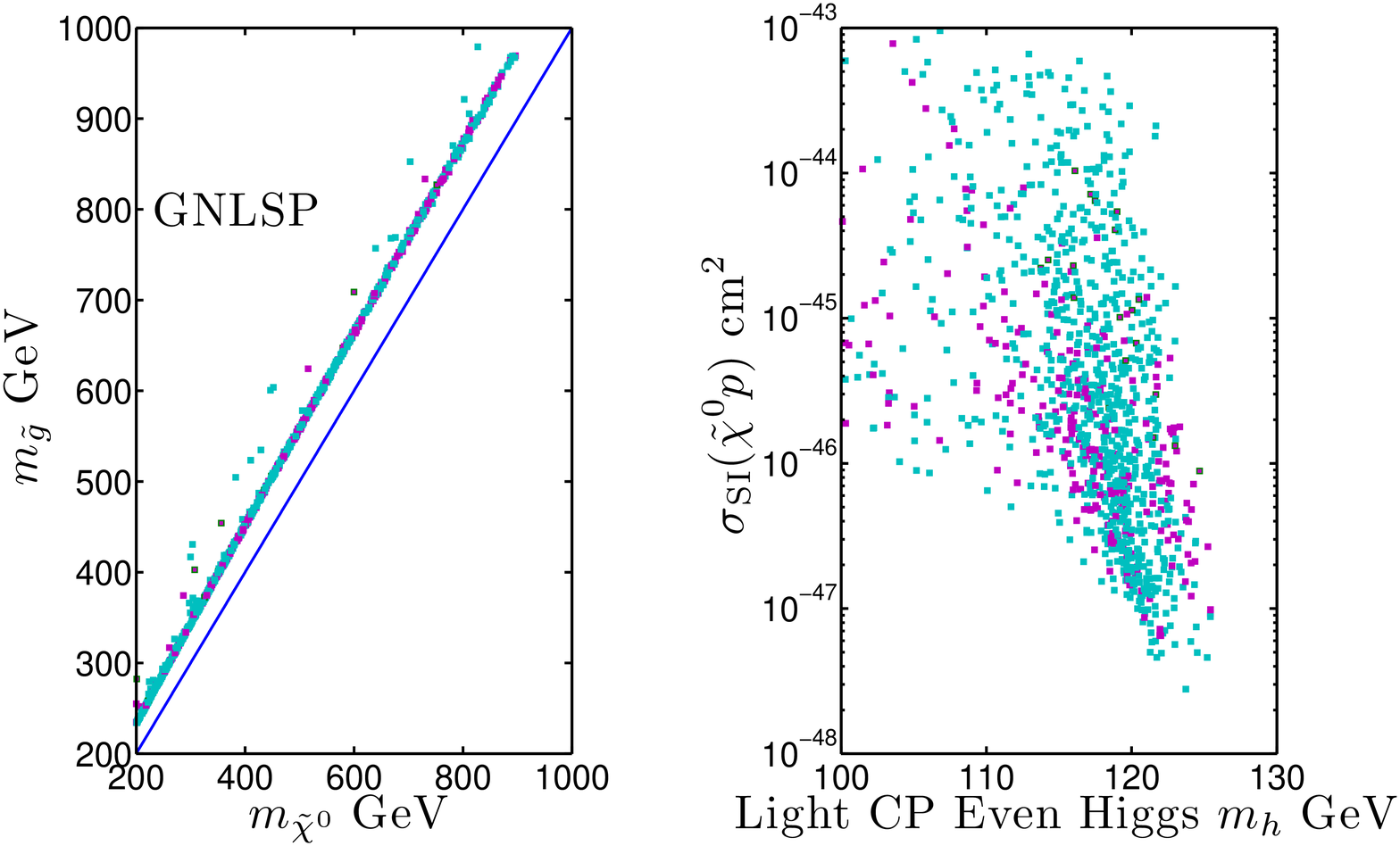}
\caption{(Color online) Left panel: An exhibition of the spin independent cross section $\sigma_{SI}$ 
as a function of the neutralino mass. 
It is seen that there are a large number of models corresponding to $\sigma_{SI}$ in the range
$10^{-46}~{\rm cm}^{2}$  and below,  which would be inaccessible in direct dark matter searches in the foreseeable future.
However, many of these models especially those with low values of $m_{\tilde\chi^0}$ (and hence of 
$m_{\tilde g}$) would be discoverable at the LHC even with low 
luminosity.  Right panel (1):  The explicit  scaling relation between $m_{\na}$ and $m_{\g}$ for the GNLSP models (also shown purely for  visual reference
is a line representing   $m_{\na} = m_{\g}$) . Right panel (2): 
An analysis of $\sigma_{SI}$  vs 
 the light Higgs boson mass illustrating that a large portion of NUSP13 (light blue) has a Higgs boson near 120 GeV while NUSP14 is 
 given in dark (magenta).
}
 \label{dark2}
  \end{center}
\end{figure*}

Many of the GNLSP model points have an LSP neutralino which is bino like.
However, there is also a significant set of models that have an LSP with large higgsino components (many of which are at high $\tan\beta$ but this is 
not exclusively so). The large $\tan \beta$ parameter points are easily spotted 
by examining the $\tan\beta$ vs $A_0/m_0$ plot in Fig.(\ref{a-tanbeta}).
The GNLSP models have important implications for the direct detection of neutralino dark matter.
An analysis of the spin independent and spin dependent cross sections in dark matter 
experiments is given in Fig.(\ref{dark2}) 
implemented with MicrOMEGAs.
Included are published limits from   the 
 ZEPLIN-III experiment\cite{Lebedenko:2008gb},  the first five-tower  CDMS data\cite{Ahmed:2008eu}
  and the  XENON 10 results \cite{Angle:2007uj}. Projected
limits (indicated by *) from CDMS and LUX are also shown \cite{Projections}.
The $\sigma_{SI}$ vs $m_{\tilde\chi^0}$ analysis shows
 some interesting results. First one finds that there are a class of GNLSP models
which are beginning to be constrained by the direct detection dark matter experiments.
These models would also produce large $\tilde g\tilde g$ production cross sections
and will be easily visible at the LHC. However, interestingly there is another 
class  of low mass neutralino (and low mass gluino) models  which have rather 
small spin independent cross sections which are outside the reach of the 
direct detection experiments in the foreseeable future. Nonetheless these models
would still lead to rather large $\tilde g\tilde g$ production cross sections and
hence will be visible at the LHC. 
Thus the LHC can detect many of the GNLSP models which are most likely
inaccessible to the direct detection dark matter experiments. 
On the flip side, for the large set of low mass GNLSP models one still has the possibility of
a very light neutralino (gluino) with a sizeable higgsino component and consequently 
a large spin independent cross section. Thus if a light gluino is indeed
indicated early on  at the LHC, it may also provide a  hint of 
the size of the dark matter  signal in direct detection of dark matter.
We note in passing that a plot of $\sigma_{SI}$ vs $m_{\tilde g}$
looks very similar to the left panel of Fig.(\ref{dark2}) as the gluino and neutralino
mass are related at the low scale via $m_{\tilde g} = (1+\Delta_{\tilde g\tilde \chi^0})m_{\na}$.

We note that in the GNLSP class of models, 
the direct 
annihilations of $\na \na$ into electron positron pairs 
is helicity suppressed. For example, we obtain 
$ \langle\sigma v\rangle_{\na \na \to e^{+}  e^{-}} \sim 5 \times 10^{-30}~ cm^3/s$ (at $v/c =.002 $) for the model of Table({\ref{GNLSP1}}) while the self 
annihilation into
$W^{+}W^{-}$ are equally small. 
Annihilations into $\tau \bar \tau$ are found to be the largest (this particular model has
$ \langle\sigma v \rangle_{\na \na \to \tau \bar \tau} \sim  2\times 10^{-27}~ cm^3/s)$. 
Thus a significant boost 
 will be needed to explain the recent cosmic ray excess\cite{CR}.
\vspace{-.5cm}

\section{GNLSP Benchmarks\label{benchmarks}}
It is useful to  give benchmarks for the three GNLSP models A, B, and C discussed in
Sec.(\ref{gaugino}). As mentioned above, we go beyond the 
perturbative calculation, 
and include the Sommerfeld enhancement of the cross section
in these benchmarks. In 
Table(\ref{benchAin}) we give benchmarks for model
GNLSP$_{\rm A}$. The benchmarks are chosen to 
exhibit a significant diversity in the input values. Some of the low lying spectrum
as well as the spin  independent cross section $\sigma_{SI}(\tilde \chi^0p)$ and the spin
dependent cross section $\sigma_{SD}(\tilde \chi^0p)$ corresponding to 
Table(\ref{benchAin}) 
are exhibited  in 
Table(\ref{benchAout}). 
The analysis of Table(\ref{benchAout})
 shows a variation
over two orders of magnitude for $\sigma_{SI}(\tilde \chi^0p)$.
As discussed already the
variation arises due  to changes in the higgsino vs the bino component of the LSP.
  Similar benchmarks for model GNLSP$_{\rm B}$ are given in 
Table(\ref{benchBin})
 and the corresponding light sparticle spectrum and the corresponding  spin  
 independent cross section $\sigma_{SI}(\tilde \chi^0p)$ and the spin dependent cross 
 section $\sigma_{SD}(\tilde \chi^0p)$ are exhibited in 
Table(\ref{benchBout}).
Finally, a similar analysis for the model  GNLSP$_{\rm C}$ is given in
 Table(\ref{benchGin}) and Table(\ref{benchGout}).   
\vspace{-.7cm}
\section{Conclusions\label{conclusions}}
\vspace{-.7cm}
In the above we have given an analysis of a class of models with nonuniversalities which lead to a  gluino as the NLSP. 
Several important observations  emerge from this analysis which have bearing on the 
observation of sparticles  at the LHC.  Perhaps the most important of  these is that  
if the gluino is the NLSP, then the $gg\to \tilde g \tilde g$ cross section at the LHC 
 dominates over all others
in the GNLSP models.
The dominance of the gluino production and  the fact that the  $\tilde g \tilde g$ production 
cross sections are large  implies that  the observation of supersymmetry  via the gluino 
production  can occur with the first data from the LHC.  It is found that the
dominant signal of the gluino NLSP model are multi-jets, tagged b-jets and missing energy
and it is possible to devise post trigger level cuts which discriminate these models above
the standard model backgrounds. Such cuts which reduce the background and enhance the signal to the
background ratio were devised and implemented in this paper. 
We note also that the inverse of the LHC process, namely $\g \g \to gg$, is 
largely responsible for the satisfaction of the relic density
when the neutralino and the gluino coannihilate. 
Further,  the analysis of GNLSP models reveals that there exists a significant region of the parameter space
in these models where  the neutralino has a large higgsino content, 
and the neutralino-proton spin independent scattering cross section is sizeable and 
can be probed with the current experimental sensitivity, and sensitivities that would be achievable 
in future experiments. However, there are also other regions of the parameter space where 
the neutralino is mostly bino like and in this case the spin independent  cross sections can 
fall well below the current experimental sensitivity, and well below the sensitivity
 that would be achievable in the near future experiments.  Interestingly, the bino cases,
though difficult to discover in dark matter experiments, can be accessible at the LHC
since the gluino mass in these models  lies within the reach of the LHC even at low luminosities.  
 Another aspect of the GNLSP models was 
 also discussed which relates to the compressed nature of the sfermion mass 
 spectrum relative to the case of universal gaugino masses.
 Here the sleptons in the first two generations could
 be almost degenerate in mass and often even heavier than their squark counterparts.
 Several benchmarks for the GNLSP models were also given to facilitate further 
 work.  It was also pointed out that a test of the GNLSP models can be done with  just 10 fb$^{-1}$ of integrated
luminosity and thus it is one of the models that can be checked with the early data at the  LHC. 
 Finally,  we emphasize once again that if the gluino is an NLSP then
 the production of gluinos will dominate all other sparticle production making the LHC 
effectively a gluino factory.

\noindent
{\em Acknowledgments}:  
This research is  supported in part by NSF grants PHY-0653342 (Stony Brook)  and PHY-0757959 (Northeastern).
The work of DF was also supported in part by Office of the Vice-Provost at Northeastern University.  

\section*{Appendix A: Gaugino mass sum rules for GUTS with nonuniversalities\label{appendixa}} 
Each of the mass ratios listed in  Table(\ref{nonunitable1}) gives two 
gaugino mass relations at the GUT  scale. Of these one is `unstable' to
the inclusion of a singlet $F$ term breaking while the other one is `stable' 
and remains valid when one includes a singlet $F$ term breaking along with the 
non-singlet breaking. Below we  list
only the `stable'  mass relations. The mass relations are labeled numerically 
(1)-(23) as in Table(\ref{nonunitable1}). They are\footnote{The stability of the gaugino mass
relations to the inclusion of singlet breaking is easily seen by noting that  the sum of the co-efficients 
of $M_1$, $M_2$ and $M_3$  in each of the mass relations in Eq.(\ref{sum}) vanishes.}
 \begin{eqnarray}
(1)+(7)+(10): -5 M_1 +3 M_2 + 2  M_3=0, ~~& (2):& M_1 + 3 M_2 - 4 M_3=0,\nonumber\\
(3):  -M_1 + 9 M_2 - 8  M_3=0,          ~~&(4): &5 M_1 +3 M_2 - 8 M_3=0,\nonumber\\
(5): - 5 M_1 + 9 M_2 - 4 M_3=0,         ~~&(9):& 5M_1+M_2 -6M_3=0,\nonumber\\
(13): ~M_1+M_2 -2M_3=0,                 ~~&(14):& - 25 M_1 +9 M_2 + 16 M_3=0,\nonumber\\
(16): -5M_1+M_2 +  4M_3=0,                 ~~&(18):& ~5M_1 + 11 M_2 -16 M_3 =0,\nonumber\\
(20): ~5 M_1+ 7 M_2 -12 M_3=0,          ~~&(23):& -5 M_1+ 9 M_2 -4 M_3=0.\nonumber
\label{sum}
\end{eqnarray}
In addition to the above, models (6), (8), (11), (12), (15), (17), (22) in Table(\ref{nonunitable1})
satisfy the
relation $M_2=M_3$, while model (19) in Table(\ref{nonunitable1})
satisfies the relation $M_1=M_3$, and model (21) in Table(\ref{nonunitable1})
satisfies the relation $M_1=M_2$.
These mass relations would  be appropriately modified at low scales
by the renormalization group evolution. Thus at the one loop level 
 the mass relations at the electroweak scale are
\beqn
(r-1)\frac{\alpha_1(0)}{\alpha_1(Q)} M_1(Q) +  \frac{\alpha_2(0)}{\alpha_2(Q)} M_2(Q)
 -r \frac{\alpha_3(0)}{\alpha_3(Q)}M_3(Q)=0,
 \eeqn
 where $r$ is the ratio as given by Table(\ref{nonunitable1}).
 Assuming one can determine with accuracy the gaugino masses, these mass relations
 can be a useful indicator of the specific $F$ type breaking
and hence of the type of nonuniversal SUGRA model one has at the GUT scale.


\section*{Appendix B: An analytic analysis of quasi-degeneracy of LSP and GNLSP and
of  the sfermion mass  compression}
Here we give a one loop analysis of how the quasi degeneracy of the LSP and of 
the GNLSP
comes about and then discuss a similar phenomenon for 
the squarks and the
sleptons for the first two generations.  We begin with the gaugino masses which
at the GUT scale obey the nonuniversality condition of Eq.(\ref{non1}) 
which we restate in more compact notation as follows: For model $i$ we have
  \beqn
   M_1^{(i)}= \gamma_1^{(i)}\mh,
     ~M_2^{(i)}= \gamma_2^{(i)}\mh,
       ~M_3^{(i)}= \gamma_3^{(i)}\mh,
       \label{non2}
  \eeqn
where
  \beqn
 \gamma_1^{(i)}= (1+a_i \alpha_i),~
 \gamma_2^{(i)} = (1+b_i \alpha_i),~
 \gamma_3^{(i)}= (1+c_i \alpha_i).
       \label{non3}
  \eeqn
With the above boundary conditions, the gaugino masses at the electroweak scale
$t=ln(M_G^2/Q^2)$, where $M_G$ is the GUT scale and $Q$ is the electroweak scale,
are given by 
\beqn
M_a^{(i)}(t)= \gamma^{(i)}_a \alpha_a(t) \mh, ~~a=1,2,3. 
\eeqn
Here a=1,2,3 correspond to the $U(1)$, $SU(2)$ and $SU(3)$ gauge groups,
and $\alpha_a(t)$ are the corresponding fine structure constants for these
groups at the electroweak scale.  For the mSUGRA case, $\gamma^{(i)}_a=1$, and
one simply has that at one loop 
\beqn
M_1(t): M_2(t): M_3(t)= \alpha_1(t):\alpha_2(t):\alpha_3(t).
\eeqn
 Using the experimental values of the
gauge coupling constants at the electroweak scale one finds the three gaugino
masses roughly  in the ratio $1:\sim 2:\sim 5-6$. In this case the gaugino masses
are split in a very significant way. However, in the presence of nonuniversalities
the ratios will be modified in a very different  way. Thus with the inclusion of the
modifications $\gamma_a$,  the three gaugino masses at the electroweak
scale will be roughly in the ratio $\gamma_1:\sim 2 \gamma_2:\sim (5-6) \gamma_3$.
It is clear then that the choice $\gamma_3/\gamma_1\sim (1/5-1/6)$ will make 
masses of gaugino 1 and of gaugino 3 almost degenerate. Of course, more realistically
there would be mixings between the gauginos and the higgsinos and the mass eigenstates
will be admixtures of these.  Thus the mass relation of the lightest neutralino and 
of the gluino will be more complicated. Still the above approximation may
roughly hold when the neutralino is mostly a Bino.\\

Nonuniversalities also enter in the masses for the squarks and for the sleptons.
For the first two generation down squarks one finds
\beqn
m^2_{\tilde d_{iL}}(t)=m_0^2+m_{di}^2
+\tilde\alpha_G [\frac{8}{3}{\tilde f}_3
+\frac{3}{2}{\tilde f}_2+ 
\frac{1}{30}{\tilde f}_1] \mh^2
+(-\frac{1}{2}+\frac{1}{3}\sin^2\theta_W) M_Z^2 \cos (2\beta),\nonumber\\
m^2_{\tilde d_{iR}}(t)=m_0^2+m_{di}^2
+\tilde\alpha_G [\frac{8}{3}{\tilde f}_3+\frac{8}{15}{\tilde f}_1] \mh^2
-\frac{1}{3}\sin^2\theta_W M_Z^2 \cos (2\beta),
\eeqn
where 
\beqn
\tilde f_a= \gamma_a^2 f_a, ~~f_a(t)= \frac{1}{\beta_a}\left(1-\frac{1}{(1+\beta_a t)^2}\right), 
\eeqn
where 
$\beta_a=b_a \tilde \alpha_a(0)$, $\tilde \alpha_a(0) =\alpha_a/4\pi$,
and $b_a=(33/5, 1, -3)$ for the gauge groups $U(1)$, $SU(2)$ and $SU(3)$. 
For the case of the first two generations of charged leptons one has 
\beqn
m^2_{\tilde e_{iL}}(t)=m_0^2+m_{ei}^2
+\tilde\alpha_G [
\frac{3}{2}{\tilde f}_2+ 
\frac{3}{10}{\tilde f}_1] \mh^2
+(-\frac{1}{2}+\sin^2\theta_W) M_Z^2 \cos (2\beta),\nonumber\\
m^2_{\tilde e_{iR}}(t)=m_0^2+m_{ei}^2
+\frac{6}{5} \tilde\alpha_G {\tilde f}_1  \mh^2
-\sin^2\theta_W M_Z^2 \cos (2\beta).
\eeqn
Using the above we find that 
\beqn
  \frac{1}{2}\left[m^2_{\tilde d_{iL}}(t)+ m^2_{\tilde d_{iR}}(t)
-(m^2_{\tilde e_{iL}}(t)+ m^2_{\tilde e_{iR}}(t))\right]
= m_{di}^2 - m_{ei}^2 + \tilde\alpha_G [\frac{8}{3}\gamma_3^2 f_3- \frac{7}{15}\gamma_1^2 f_1]. 
\label{diff}
\eeqn
It is now easily seen that the last brace in Eq.(\ref{diff}) can vanish or even
turn negative by appropriate choice of $\gamma_3$ vs $\gamma_1$. The above
situation leads to a near equality of the squark and of the slepton masses
discussed in the text of the paper.  
The $\mu$ parameter also has a strong dependence on
nonuniversalities. This is seen by examining the relation that determines 
$\mu^2$, i.e.,
\beq 
\mu^2=(m_{H_1}^2-m_{H_2}^2\tan\beta^2)(\tan\beta^2-1)^{-1}
-\frac{1}{2}M_Z^2+ \Delta \mu^2,
\eeq
where  $\Delta\mu^2$ is the loop correction. Thus $m_{H_1}$, and
 $m_{H_2}$ are sensitive to nonuniversalities since 
\beq
m_{H_1}^2=m_0^2+\tilde\alpha_G \left(\frac{3}{2} {\tilde f}_2(t) +
\frac{3}{10} {\tilde f}_1(t)\right) \mh^2,
\eeq
where the nonuniversalities enter via the ${\tilde f}$ functions. Similarly 
$m_{H_2}$ is given by 
\beq
  m_{H_2}^2=\mh^2{\tilde e}(t)
+A_0m_0 \mh {\tilde f}(t)+m_0^2(h(t)-k(t)A_0^2),
\eeq	 
where the tilde functions $\tilde e$ and $\tilde f$ are modified due to
nonuniversality while the functions  $h(t)$ and $k(t)$ are not 
unaffected (for definitions of these functions see the first paper 
of \cite{nonuni2} and the references therein which also gives a more detailed
discussion of this topic).
The dependence on nonuniversalties is more complicated in
this case because of the coupling with the top quark.
Specifically, one can derive the following relation which gives the 
explicit dependence on nonuniversalities
\beqn
   \frac{\partial \mu^2}{\partial \gamma_a}
   =(t^2-1)^{-1}(\mh^2g_a'-t^2(\mh^2e_a'+A_0m_0\mh f_a'))
   + \frac{\partial\Delta \mu^2}{\partial \gamma_a},
  \eeqn   
  where
$ g_a'=\frac{\partial \tilde g}{\partial \gamma_a}$,  
$\tilde g=\tilde \alpha_G(\frac{3}{2}\tilde f_2+\frac{3}{10}\tilde f_1)$,
$e'_a=\frac{\partial \tilde e}{\partial \gamma_a}$,
and $f_a'=\frac{\partial \tilde f}{\partial \gamma_a}$.
One can make a  semi-quantitative estimate of the dependence of $\mu^2$ 
on nonuniversalities from above. We note, however, that 
$\mu$ does not enter sensitively in the sum rule for the first two
generation of squarks and sleptons and thus an estimate of the compression of the sfermion
spectrum in the first two generations can be made without estimate of the $\mu$ parameter.

\section*{Appendix C: Sample GNLSP benchmarks\label{appendixc}} 
We provide here sample model points for the GNLSP class of models. 
Each model point obeys experimental constraints as discussed in the text. We use
here SuSpect 2.41 coupled to MicrOMEGAs (MO) 2.2.CPC  along with an independant code
which agrees with MO at the pertuabtive level but accounts for the non-perturbative effects discussed in the text.
 Similar model points
may be obtained with other spectrum calculators coupled to MO [see, for example, 
Table(\ref{GNLSPcompare}) where one such comparison is given, which, however, is 
 only at the 
perturbative level].
\begin{table*}[h]	
\vspace{2cm}																			
\begin{center}																			
\begin{tabular}{|c||c|c|c|c|c|c|c|}																					\hline\hline		
{\bf GNSLP}	& 		$m_0$ (GeV)	&	$ m_{1/2}$ (GeV)	&	$A_0$ (GeV)	&	$ \tan \beta$	&	$\delta_2$	&	$\delta_3$\\	
\hline\hline																			
$\rm GNSLP_{ A1 }$	&	2949	&	692	&	3658	&	35	&	0.566	&	-0.847	\cr
$\rm GNSLP_{ A2 }$	&	2706	&	783	&	4408	&	37	&	0.560	&	-0.839	\cr
\hline
$\rm GNSLP_{ A3 }$	&	2529	&	946	&	3873	&	41	&	0.560	&	-0.837	\cr
$\rm GNSLP_{ A4 }$	&	2967	&	910	&	5114	&	27	&	0.557	&	-0.834	\cr
\hline
$\rm GNSLP_{ A5 }$	&	2574	&	1058	&	4197	&	42	&	0.557	&	-0.833	\cr
$\rm GNSLP_{ A6 }$	&	2821	&	1019	&	5050	&	20	&	0.554	&	-0.830	\cr
\hline
$\rm GNSLP_{ A7 }$	&	3008	&	1252	&	-3241	&	27	&	0.558	&	-0.836	\cr
$\rm GNSLP_{ A8 }$	&	2746	&	1265	&	5186	&	17	&	0.551	&	-0.824	\cr
\hline\hline
\end{tabular}	
\caption{ {\bf \a ~benchmarks:}  These models produce the correct relic density
with SuSpect 2.41 coupled to MO 2.2.CPC including the non-perturbative corrections . 
Benchmarks here have $(\Omega h^2)_{\chi^0} \in (0.100,0.130)$. Here $m_t({\rm pole})/\GeV$ =170.9 throughout.} 															\label{benchAin}				
\end{center}
\vspace{2cm}																			
\begin{center}		
\begin{tabular}{|c||c|c|c|c|c|c|c|c|c|}																	
\hline\hline																			
{\bf GNSLP} 	&	$m_{h}$	&	$m_{\na}$   	&	$m_{\g}$   	&	$m_{\cha}$	&	$m_{\ta}$	&	 $m_{A}| {\tilde h}^{\rm Frac}_{1,2}$  		&	$\sigma_{SI}(\na p)$ 	&	$\sigma_{SD}(\na p)$	\\
Model 	&	(GeV)	&	 (GeV)	&	 (GeV)	&	 (GeV)	& (GeV)	&	$(\rm GeV)|-$		&	(pb)	&	(pb)	\\
\hline\hline																		
$\rm GNSLP_{ A1 }$	&	117	&	285	&	343	&	343	&	1560	&	2130	$|$	0.241	&	3.0$\times 10^{-8}$	&8.0$\times 10^{-5}$	\cr
$\rm GNSLP_{ A2 }$	&	117	&	337	&	387	&	869	&	1225	&	1952	$|$	0.004	&	2.7$\times 10^{-10}$	&	2.8$\times 10^{-7}$	\cr
\hline
$\rm GNSLP_{ A3 }$	&	116	&	398	&	456	&	480	&	1190	&	1540	$|$	0.135	&	2.3$\times 10^{-8}$	&	2.4$\times 10^{-5}$\cr
$\rm GNSLP_{ A4 }$	&	117	&	399	&	454	&	1080	&	1276	&	2724	$|$	0.003	&	1.3$\times 10^{-10}$	&	1.1$\times 10^{-7}$	\cr
\hline
$\rm GNSLP_{ A5 }$	&	116	&	447	&	510	&	531	&	1161	&	1514	$|$	0.131	&	2.2$\times 10^{-8}$		&	1.9$\times 10^{-5}$	\cr
$\rm GNSLP_{ A6 }$	&	117	&	448	&	507	&	1064	&	1149	&	2886	$|$	0.003	&	2.1$\times 10^{-10}$	&	1.4$\times 10^{-7}$	\cr
\hline
$\rm GNSLP_{ A7 }$	&	120	&	551	&	618	&	645	&	1332	&	2718	$|$	0.108	&	1.3$\times 10^{-8}$		&	1.0$\times 10^{-5}$	\cr
$\rm GNSLP_{ A8 }$	&	116	&	557	&	624	&	970	&	1032	&	2960	$|$	0.007	&	6.7$\times 10^{-10}$		&	3.3$\times 10^{-7}$	\cr
				\hline\hline
\end{tabular}	
\caption{ {\bf \a ~properties:}
 The variation in the spin
independent cross section over two orders of magnitude arises due to variations in the higgsino vs the bino 
component of the LSP. The higgsino fraction has been defined by 
 ${\tilde h}^{\rm Frac}_{1,2} = |\gamma|^2+|\delta|^2$, where the normalized LSP mass eigenstate is $\tilde \chi^0=\alpha \tilde b
 + \beta \tilde w + \gamma \tilde  h_1 +\delta \tilde h_2$ as in the notation defined in Sec.(\ref{relic}).} 
\label{benchAout}					
\end{center}	
\end{table*}		

\clearpage

	
\begin{table*}[h]	
\vspace{2cm}															
\begin{center}																			
\begin{tabular}{|c||c|c|c|c|c|c|c|}																					
\hline\hline															
{\bf GNSLP}	&	$m_0$ (GeV)	&	$ m_{1/2}$ (GeV)	&	$A_0$ (GeV)	&	$ \tan \beta$	&	$\delta_2$	&	$\delta_3$\\	
\hline\hline																														
$\rm GNSLP_{ B1 }$	&	2421	&	736	&	3414	&	51	&	0.000	&	-0.841	\cr
$\rm GNSLP_{ B2 }$	&	3406	&	734	&	4655	&	35	&	0.000	&	-0.848	\cr
\hline
$\rm GNSLP_{ B3 }$	&	2890	&	945	&	-2977	&	47	&	0.000	&	-0.844	\cr
$\rm GNSLP_{ B4 }$	&	3772	&	988	&	5894	&	46	&	0.000	&	-0.837	\cr
\hline
$\rm GNSLP_{ B5 }$	&	2857	&	1158	&	-2631	&	24	&	0.000	&	-0.842	\cr
$\rm GNSLP_{ B6 }$	&	2943	&	1142	&	5006	&	12	&	0.000	&	-0.831	\cr
\hline
$\rm GNSLP_{ B7 }$	&	3188	&	1376	&	-2479	&	7	&	0.000	&	-0.837	\cr
$\rm GNSLP_{ B8 }$	&	2659	&	1380	&	5028	&	37	&	0.000	&	-0.825	\cr
\hline\hline		
\end{tabular}	
\caption{ {\bf \b ~benchmarks:} 
As in Table(\ref{benchAin}) the displayed models produce the correct relic density.
 Benchmarks here have $(\Omega h^2)_{\chi^0}  \in (0.100,0.120)$.}												
\label{benchBin}	
\end{center}
\vspace{2cm}	
\begin{center}																	
\begin{tabular}{|l||c|c|c|c|c|c|c|c|l|}																			
\hline\hline																			
{\bf GNSLP} 	&	$m_{h}$	&	$m_{\na}$   	&	$m_{\g}$   	&	$m_{\cha}$	&	$m_{\ta}$	&	 $m_{A}| {\tilde h}^{\rm Frac}_{1,2}$ 		&	$\sigma_{SI}(\na p)$ 	&	$\sigma_{SD}(\na p)$	\\
Model 	&	(GeV)	&	 (GeV)	&	 (GeV)	&	 (GeV)	& (GeV)	&	$(\rm GeV)|-$		&	(pb)	&	(pb)	\\
\hline\hline
$\rm GNSLP_{ B1 }$	&	116	&	313	&	364	&	557	&	1223	&	409	$|$	0.009	&	2.4$\times 10^{-8}$	&	1.2$\times 10^{-6}$	\cr
$\rm GNSLP_{ B2 }$	&	119	&	316	&	364	&	576	&	1734	&	2520	$|$	0.003	&	1.4$\times 10^{-10}$	&	2.0$\times 10^{-7}$	\cr
\hline
$\rm GNSLP_{ B3 }$	&	118	&	417	&	484	&	582	&	1507	&	665	$|$	0.035	&	2.0$\times 10^{-8}$&	5.2$\times 10^{-6}$\cr
$\rm GNSLP_{ B4 }$	&	119	&	428	&	488	&	779	&	1818	&	1756	$|$	0.002	&	1.0$\times 10^{-10}$	&	5.6$\times 10^{-8}$	\cr
\hline
$\rm GNSLP_{ B5 }$	&	119	&	502	&	566	&	583	&	1475	&	2610	$|$	0.140	&	1.8$\times 10^{-8}$	&	1.7$\times 10^{-5}$	\cr
$\rm GNSLP_{ B6 }$	&	117	&	502	&	565	&	948	&	1299	&	3177	$|$	0.002	&	1.7$\times 10^{-10}$	&	6.8$\times 10^{-8}$	\cr
\hline
$\rm GNSLP_{ B7 }$	&	117	&	601	&	672	&	701	&	1676	&	3382	$|$	0.113	&	1.8$\times 10^{-8}$	&	7.4$\times 10^{-6}$	\cr
$\rm GNSLP_{ B8 }$	&	116	&	598	&	669	&	1051	&	1121	&	2031	$|$	0.004	&	4.3$\times 10^{-10}$	&	1.6$\times 10^{-7}$\cr
\hline\hline																			
\end{tabular}	
\caption{ {\bf \b ~properties:}
The table gives an analysis similar to that of Table(\ref{benchAout}) for $\rm GNLSP_{B}$ models.}		
\label{benchBout}		
\end{center}\end{table*}	




\begin{table*}[htbp]																			
\begin{center}																			
\begin{tabular}{|l||c|c|c|c|c|c|l|}																					
\hline\hline															
{\bf GNSLP} 	 	&	$m_0$ (GeV)	&	$ m_{1/2}$ (GeV)	&	$A_0$ (GeV)	&	$ \tan \beta$	&	$\delta_2$	&	$\delta_3$\\	
\hline\hline																			
$\rm GNSLP_{ C1 }$	&	1604	&	450	&	2035	&	49	&	-0.317	&	-0.852	\cr
$\rm GNSLP_{ C2 }$	&	2119	&	696	&	2860	&	44	&	0.291	&	-0.845	\cr
$\rm GNSLP_{ C3 }$	&	2443	&	948	&	2823	&	11	&	-0.227	&	-0.842	\cr
$\rm GNSLP_{ C4 }$	&	3850	&	1111	&	4388	&	9	&	0.209	&	-0.840	\cr
$\rm GNSLP_{ C5 }$	&	2599	&	1270	&	3656	&	14	&	-0.009	&	-0.833	\cr
$\rm GNSLP_{ C6 }$	&	2458	&	1479	&	5045	&	35	&	0.462	&	-0.823	\cr
\hline
$\rm GNSLP_{ C7 }$	&	2087	&	453	&	2359	&	21	&	0.292	&	-0.862	\cr
$\rm GNSLP_{ C8 }$	&	1958	&	674	&	2950	&	22	&	0.299	&	-0.843	\cr
$\rm GNSLP_{ C9 }$	&	3874	&	1098	&	4455	&	8	&	-0.018	&	-0.839	\cr
$\rm GNSLP_{ C10 }$	&	2543	&	1337	&	4188	&	49	&	-0.147	&	-0.826	\cr
$\rm GNSLP_{ C11 }$	&	3288	&	1431	&	5879	&	15	&	0.703	&	-0.822	\cr
$\rm GNSLP_{ C12 }$	&	3942	&	1755	&	5995	&	36	&	0.080	&	-0.825	\cr
\hline\hline
\end{tabular}
\caption{ {\bf \c ~benchmarks:} A sample of benchmarks in \c in a random distribution in $\delta_2$ and $\delta_3$.  Many of the models  listed above have a substantial higgsino component and
part of the parameter space would be accessible to current and future experiments for the direct 
detection of dark matter. Benchmarks here have  $(\Omega h^2)_{\chi^0} \in (0.100,0.138)$.}
			\label{benchGin}
\end{center}																			
\begin{center}																			
\begin{tabular}{|l||c|c|c|c|c|c|c|c|l|}																	
\hline\hline																			
{\bf GNSLP} 	&	$m_{h}$	&	$m_{\na}$   	&	$m_{\g}$   	&	$m_{\cha}$	&	$m_{\ta}$	 & $m_{A}| {\tilde h}^{\rm Frac}_{1,2}$  		&	$\sigma_{SI}(\na p)$ 	&	$\sigma_{SD}(\na p)$	\\
Model 	&	(GeV)	&	 (GeV)	&	 (GeV)	&	 (GeV)	& (GeV)	&	$(\rm GeV)|-$		&	(pb)	&	(pb)	\\
\hline\hline
$\rm GNSLP_{ C1 }$	&	113	&	185	&	227	&	228	&	830	&	516	$|$	0.022	&	3.6$\times 10^{-8}$	&	9.0$\times 10^{-6}$\cr
$\rm GNSLP_{ C2 }$	&	115	&	291	&	338	&	423	&	1073	&	1065	$|$	0.056	&	1.4$\times 10^{-8}$	&	1.6$\times 10^{-5}$\cr
$\rm GNSLP_{ C3 }$	&	115	&	404	&	461	&	508	&	1315	&	2488	$|$	0.075	&	1.3$\times 10^{-8}$	&	1.2$\times 10^{-5}$\cr
$\rm GNSLP_{ C4 }$	&	117	&	481	&	546	&	587	&	2078	&	3939	$|$	0.098	&	1.4$\times 10^{-8}$	&	1.0$\times 10^{-5}$	\cr
$\rm GNSLP_{ C5 }$	&	116	&	542	&	610	&	609	&	1320	&	2677	$|$	0.200	&	2.9$\times 10^{-8}$&	1.9$\times 10^{-5}$	\cr
$\rm GNSLP_{ C6 }$	&	115	&	632	&	704	&	716	&	830	&	2042	$|$	0.130	&	2.0$\times 10^{-8}$	&	9.7$\times 10^{-6}$	\cr
\hline
$\rm GNSLP_{ C7 }$	&	115	&	189	&	223	&	333	&	1114	&	1938	$|$	0.047	&	5.1$\times 10^{-9}$	&	2.2$\times 10^{-5}$	\cr
$\rm GNSLP_{ C8 }$	&	114	&	289	&	333	&	581	&	924	&	1887	$|$	0.012	&	1.2$\times 10^{-9}$	&	1.8$\times 10^{-6}$\cr
$\rm GNSLP_{ C9 }$	&	117	&	483	&	546	&	816	&	2088	&	3996	$|$	0.010	&	1.1$\times 10^{-9}$	&	6.0$\times 10^{-7}$	\cr
$\rm GNSLP_{ C10 }$	&	116	&	576	&	648	&	797	&	1242	&	956	$|$	0.018	&	5.4$\times 10^{-9}$	&	1.3$\times 10^{-6}$	\cr
$\rm GNSLP_{ C11 }$	&	117	&	634	&	708	&	909	&	1310	&	3558	$|$	0.015	&	1.7$\times 10^{-9}$	&	8.0$\times 10^{-7}$	\cr
$\rm GNSLP_{ C12 }$	&	119	&	769	&	849	&	948	&	1965	&	2909	$|$	0.031	&	3.7$\times 10^{-9}$	&	1.6$\times 10^{-6}$	\cr
\hline\hline
\end{tabular}	
\caption{{\bf \c ~properties:}
A display of a part of the sparticle mass spectrum consisting of light Higgs and CP odd Higgs masses, and
masses of the LSP, NLSP, light chargino, and light stop along with spin independent and dependent 
cross sections. Models shown here include those with 
 higgsino like LSPs and as well as  those with mixed higgsino and bino LSPs, and LSPs
which are mostly bino. The horizontal line emphasizes separation of the $\sigma_{SI}/\pb$ from $O(10^{-8})$ to $O(10^{-9})$ . }			
\label{benchGout}
\end{center}	
\label{tab11}																		
\end{table*}	

\clearpage

\end{document}